\documentclass[a4paper,usenatbib]{mn2e}

\usepackage[totalwidth=500pt,totalheight=680pt]{geometry}
\usepackage{graphicx}
\usepackage{tabulary}
\usepackage{epsfig}
\usepackage{amssymb}
\usepackage{amsmath}
\usepackage{multirow}
\usepackage{appendix}
\usepackage{natbib}
\usepackage{paralist}
\usepackage{color}
\usepackage{subfigure}
\pdfoutput=1

\newcommand{\beq}[1]{\begin{equation}\label{#1}}
\newcommand{\eeq}{\end{equation}}

\newcommand{\Ms}{M_{\star}}

\renewcommand{\bar}{\overline}

\newcommand{\EXP}[1]{\times 10^{#1}}
\newcommand{\Tau}{\mathcal{T}}
\newcommand{\Ttid}{\Tau_{\rm tid}}
\newcommand{\Ttri}{\Tau_{\rm tri}}
\newcommand{\Sgn}{{\rm Sgn}}
\newcommand{\Mp}{M_{\rm p}}
\newcommand{\Rp}{R_{\rm p}}
\newcommand{\avTtid}{\langle\Ttid\rangle}

\title[Spin-Orbit Evolution of the GJ 667C System] {The Spin-Orbit
  Evolution of GJ 667C System: \\ The Effect of Composition and Other
  Planet's Perturbations} \author[Cuartas-Restrepo, Melita, Zuluaga,
  Portilla-Revelo, Sucerquia, Miloni] {P.A.~Cuartas-Restrepo$^1$\thanks{E-mail:
    pablo.cuartas@udea.edu.co}, M.~Melita$^2$, J.I.~Zuluaga$^1$,
  \newauthor B.~Portilla-Revelo$^1$, M.~Sucerquia$^1$,
  O.~Miloni$^3$. \\ \\ $^1$FACom - SEAP - Instituto de F\'{\i}sica -
  FCEN, Universidad de Antioquia, Calle 70 No. 52-21, Medell\'{\i}n,
  Colombia.\\ $^2$Instituto de Astronom\'ia y F\'isica del Espacio -
  IAFE, Buenos Aires, Argentina. \\ $^3$FCAG - Universidad Nacional de
  La Plata, La Plata, Argentina.}  \date{Released 2002 Xxxxx XX}

\pagerange{\pageref{firstpage}--\pageref{lastpage}} \pubyear{2002}

\def\LaTeX{L\kern-.36em\raise.3ex\hbox{a}\kern-.15em
    T\kern-.1667em\lower.7ex\hbox{E}\kern-.125emX}

\begin{document}

\label{firstpage}

\maketitle

\begin{abstract}
Potentially habitable planets within the habitable zone of M-dwarfs
are affected by tidal interaction. We studied the tidal evolution in GJ 667C 
using a numerical code we call TIDEV. We reviewed the problem of the
dynamical evolution focusing on the effects that a rheological
treatment, different compositions and the inclusion of
orbital perturbations, have on the spin-down time and the probability
to be trapped in a low spin-orbit resonance.  Composition have a noticiable 
effect on the spin-down time, changing, in some cases, by almost a
factor of 2 with respect to the value estimated for a reference
Earth-like model. We calculated the time to reach a low resonance
value (3:2) for the configuration of 6 planets. Capture probabilities are
affected when assuming different compositions and eccentricities
variations.  We chose planets b and c to evaluate the probabilities of
capture in resonances below 5:2 for two compositions:
Earth-like and Waterworld planets. We found that perturbations, 
although having a secular effect on eccentricities,
have a low impact on capture probabilities and nothing on spin-down
times. The implications of the eccentricity variations and actual 
habitability of the GJ 667C system are discussed.
\end{abstract}



\section{Introduction}
\label{sec:introduction}

Finding a habitable Earth-like planet has become the holy grail of
exoplanetary research.  Planetary habitability is a complex
property constrained by factors ranging from stellar insolation and the
characteristics of planetary atmosphere to interior structure,
magnetic field strength and rate and direction of planetary rotation
\citep{Lammer10}.  

The evaluation of a few of these factors on true
exoplanets is relatively straight forward (e.g. stellar insolation or
planetary gravity), but others (e.g. planetary magnetic fields and
rotation), remain elusive and are still a matter of investigation
\citep{Zuluaga12, Zuluaga13}.

In the case of low mass stars (M-dwarfs or dMs), the Habitable Zone
(HZ) is close to the star. Under this condition, tidal
interactions can be modeled to estimate the final rotational state of
nearby planets. From there, it is possible to constrain other
properties that depend on rotation, which could affect habitability,
like intrinsic magnetic fields \citep{Zuluaga12, Zuluaga13}.

Spin-orbit evolution of Earth-like exoplanets and super-Earths, like
planets in the GJ 667C system, have been a matter of research in the
last couple of years \citep{Heller11, Rodriguez12, Callegari12,
  Anglada13}.  Most of these approaches have used a traditional tidal
formalism where torques are frequency independent, neglecting the
complexity of planetary rheology \citep{Correia08, FerrazMello08,
  Leconte10}.
  
The dependence on the rheology remains an open
discussion. Recently, \citet{FM13} and \citet{Correia14} reviewed the
problem of the tidal torque; they found that for rocky bodies, 
independent of the eccentricity, the
stationary rotation is close to the resonance.  The second work by
\citet{Correia14} assumes a Maxwell's classic viscoelastic rheology to
calculate the body deformation and applies their model to already
discovered super-Earths. For this kind of planets, they found that the
rotation stops temporarily in synchronous resonances during the
spin-orbit evolution. Eventually, these stationary stages are crossed as
long as the eccentricity decreases. They suggest that in the presence
of other planets in the system, high order resonances could remain
during all the planetary evolution. They conclude that close-in 
Earth-like planets could be captured in non-synchronous spin-orbit
resonances.
  
On the other hand, another approach to the problem was developed and
applied successfully by Efroimsky, Makarov \& Williams in their works 
\citep{EW09, ME12, E12a, M12, WE12, MB14,
  Makarov15}. This model, based on the seed work by \citet{Darwin79}
and developed later by \citet{Kaula64}, treats realistically the
dependency on rheology, and hence, on frequency of the tidal torque.
  
In this work we applied the {\it Efroimsky-Makarov-Williams} treatment
(hereafter EMW) included numerically in a package developed by
ourselves that we call {\it Tidal Evolution Package - TIDEV}, to
analize the tidal interactions and spin-orbit evolution in the GJ 667C
planetary system. Although \citet{MB14} recently applied the same
formalism to study this planetary system, their approach neglected the
possible existence of other planets around GJ 667C, announced by
\citet{Anglada13}. We analyzed three possible configurations for the
planetary system, including the possible configuration of 2 and 3
planets, recently announced by \citet{Feroz14}. For our study purposes, 
we used the main physical and orbital parameters of the system that has
been updated by \citet{Anglada13} (see Table
\ref{tab:SystemProperties}).

Regarding numerical codes devoted to the evolution 
of tides in compact planetary systems, the recent work of 
\cite{Bolmont15} proposes a tool based on the well-known code 
\textit {Mercury}, to which they have added the equations of tidal 
torque and other aspects as the relativistic effects and the deformation 
of the bodies. They have named their code \textit {Mercury-T}. 
This work includes a classic 
treatment of the tidal evolution problem using a model of constant delay 
(constant lag) without averaged equations, which allows 
them to calculate the crossings through resonances and the 
probabilities of tidal locking. Nevertheless, they do not include the 
gravitational interaction between other possible members of the system. 
In Section \ref {sec:Model}, we do a more detailed comparison with our 
own code of tidal evolution.

Our aim in this work is to study the impact that planetary composition could
have on the tidal evolution of a multiplanetary system and
its final rotational state, regardless of the number of members of the
system. In the case of GJ 667C, being a dynamically-packed system,
the effect of orbital perturbations could also be important when
studying the tidal interaction and its evolution. Here, we present and
apply the EMW approach and include the effect of secular perturbations
in the integration of tidal evolution equations 
using TIDEV, that includes the tidal equations 
proposed mainly in the works by \cite{EW09}, 
\cite{E12a}, \cite{ME12} and \cite{EM13}.

This paper is organized as follows: Section \ref{sec:GJ 667C} 
describes the main properties of the GJ 667C system
including a complete analysis of the secular evolution of three
different configurations: 2, 3 and 6 planets.  In section
\ref{sec:TidalEvolution} we describe the basic theory and formalism of
tidal torque proposed by EMW. Section \ref{sec:Model} explains 
our theoretical model, our numerical experiments and the
analysis of the effect of the other members of the planetary system
over the dynamical evolution of planets $b$ and $c$. Section
\ref{sec:Results} shows our main results.  Finally, in
Section \ref{sec:Conc} we discuss the results and talk about the
actual habitability of GJ 667C.

\begin{table*}
\caption{Astrocentric orbital elements and bulk properties for the 
configuration of 6 planets in GJ 667C System used in this work. The main
  values for period, semimajor axe and mass are adopted from
  \citet{Anglada13}. Values for mean, min and max excentricities are
  results from our model of the dynamical evolution of the system.
  Radius, in the last column, is estimated by using an interior structure
  model assuming two extreme compositions: an Earth-like and a
  Waterworld-like planet. For composition parameters see Table
  \ref{tab:InputParameters}. }
 \label{tab:SystemProperties}
 \begin{tabular}{cccccccccc}
  \hline \textbf{Planet} & \textbf{$P$(d)} & \textbf{$a$(AU)} & \textbf{$\bar{e}$} & \textbf{$e_{\rm min}$} & \textbf{$e_{\rm max}$}
  & $\omega$ & $M_0$ & \textbf{Msin\textit{i}($M_{\oplus}$)} & \textbf{R}($R_{\oplus}$) \\ 
  \hline\hline 
  b & 7.2006 & 0.05043 & 0.08118 & 0.04152 & 0.11711 & 4.97 & 209.18 & 5.94 & 1.6 - 1.9 \\ 
  c & 28.1231 & 0.12507 & 0.02248 & 0.00006 & 0.04806 & 101.38 & 154.86 & 3.86 & 1.4 - 1.6 \\ 
  f & 39.0817 & 0.15575 & 0.04220 & 0.00023 & 0.07449 & 77.73 & 339.39 & 1.94 & 1.2 - 1.3 \\ 
  e & 62.2657 & 0.21246 & 0.02297 & 0.00002 & 0.04809 & 317.43 & 11.32 & 2.68 & 1.3 - 1.4 \\ 
  d & 92.0926 & 0.27580 & 0.04244 & 0.00016 & 0.08145 & 126.05 & 243.43 & 5.21 & 1.6 - 1.8 \\ 
  g & 251.519 & 0.53888 & 0.10533 & 0.09696 & 0.11349 & 339.48 & 196.53 & 4.41 & 1.5 - 1.7 \\ 
  \hline
\end{tabular}
\end{table*}

\begin{figure}  
 \subfigure{\includegraphics[width=62mm, angle=-90]{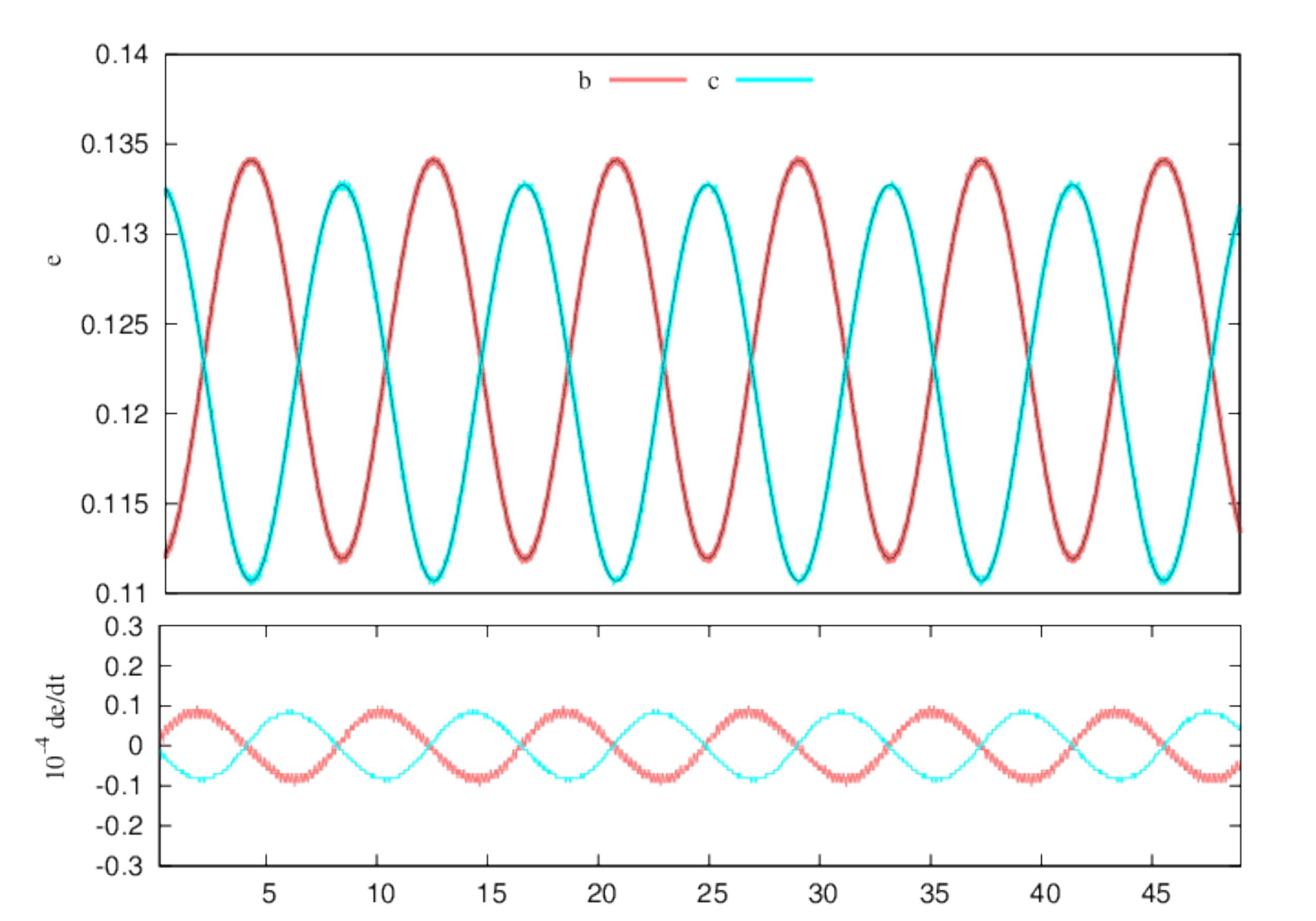}}
 \subfigure{\includegraphics[width=62mm, angle=-90]{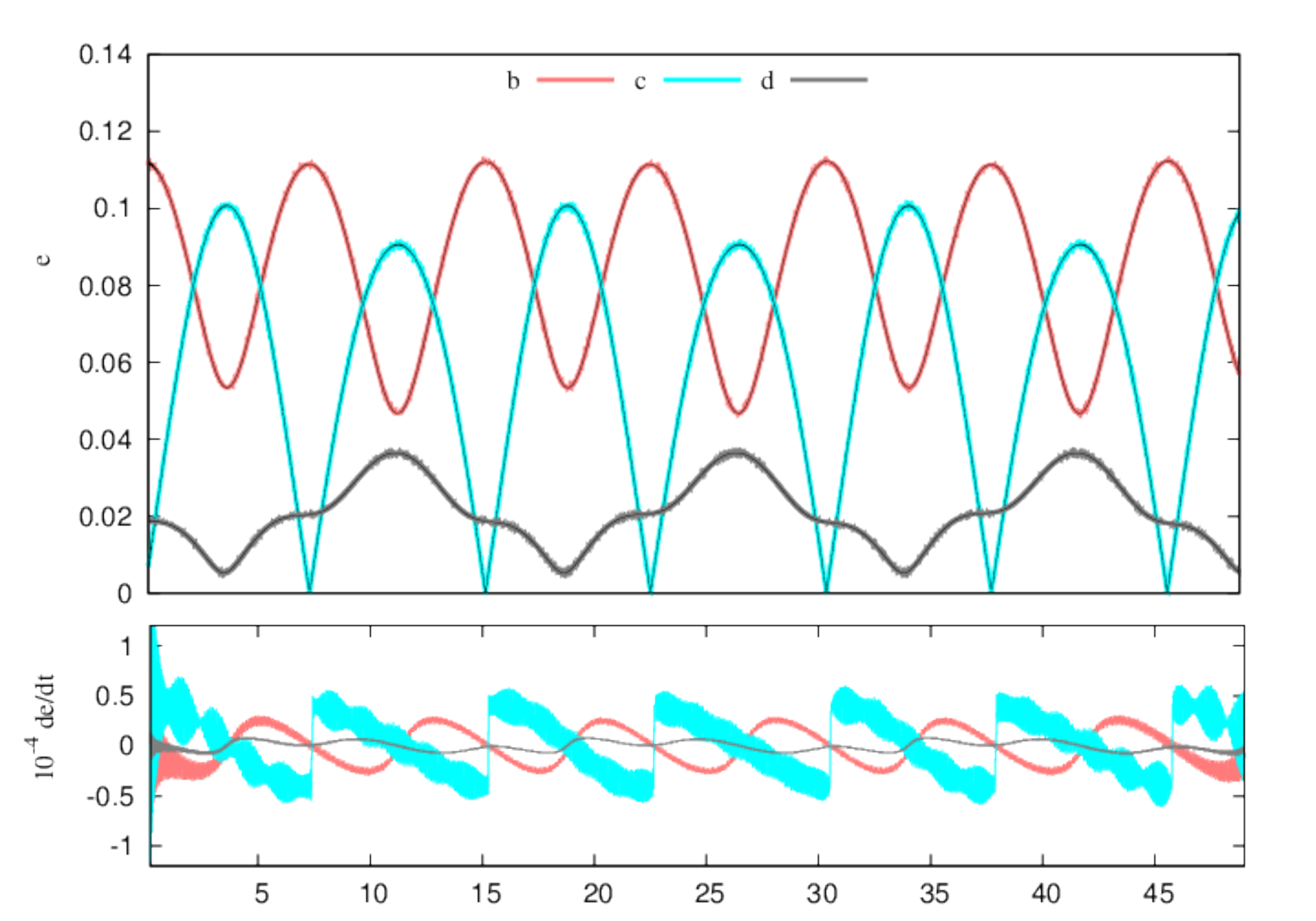}}
 \subfigure{\includegraphics[width=62mm, angle=-90]{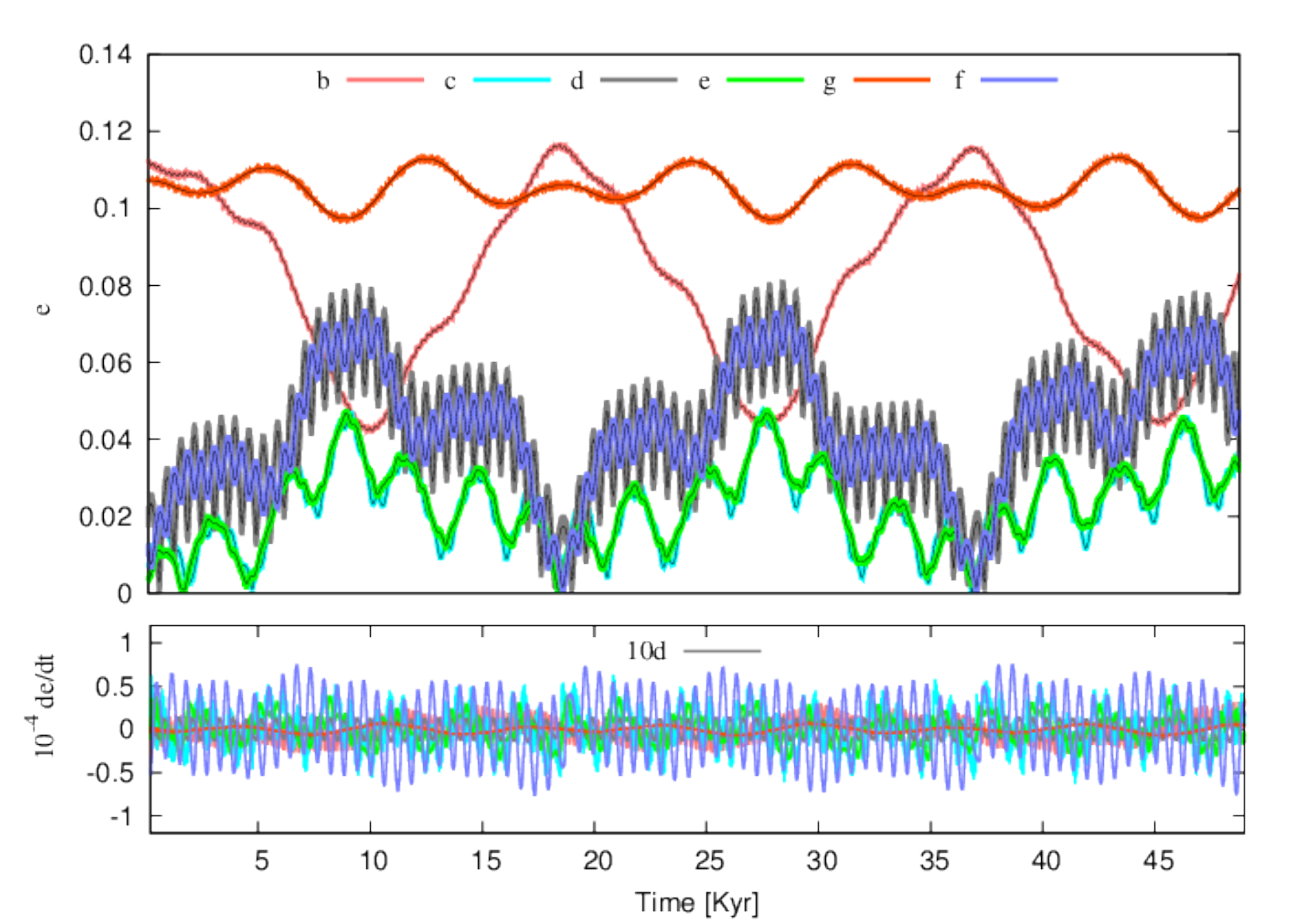}}
 \caption{Evolution of eccentricity for GJ 667C system. Three
   different configurations are showing: 2, 3 and 6
   planets. The colored curves corresponds to the
     evolution of eccentricity obtained by numerical integration for
     each planet. The inner curve in black, inside each colored curve,
     corresponds to the eccentricity evolution reconstructed
     semi-analytically for each planet.}
\label{fig:SystemSimulation}
\end{figure}
  
\section{The GJ 667C System}
\label{sec:GJ 667C}

GJ 667C is a M1.5V, 0.33 $M_{\odot}$ member of a triple stellar
system, located at 6.8 pc from the Earth in the direction of
Scorpius. The star is in a wide orbit $\sim 230$ AU away from the
center of mass of two close packed companions, GJ 667A and GJ 667B
\citep{Anglada13}. Spectroscopic studies suggest that the main binary
members (AB) are main sequence stars with an estimated age larger than
2 Gyr \citep{Cayrel81}. Independently, the present-day measured
rotational period of GJ 667C ($\sim 100$ days) and its signals of
chromospheric activity points out to an age $\sim$8 Gyr \citep{Anglada13}.

The GJ 667 system has recently been a cause of debate because of the actual
number of planets orbiting the C star. Lately, It has been shown that
this planetary system has at least two planets, one of them (planet c)
within the HZ \citep{Anglada12, Anglada13, Feroz14}. The
configuration proposed by \citet{Anglada13} is 6 planets closely
packed in orbits with semi-major axes between 0.05 and 0.5 AU. Table
\ref{tab:SystemProperties} summarizes the basic physical and orbital
properties of the system as reported by \citet{Anglada13}. In their
work, they propose the possible existence of a seventh planet
with an orbit among the already announced members. The observed mass
of the planets, between 1.94-5.94 $M_{\oplus}$, locate them in the
mass range of super-Earths. However, and according to Kepler's most 
recent results, there is a significant chance that at least the
more massive members of the system could actually be mini Neptunes,
volatile rich planets \citep{Benneke13}.  Planetary formation theories
suggest that this kind of planets could be water rich and hence, if
solid, true Waterworlds \citep{Kaltenegger13}.

On the other hand, a more recent work by \citet{Feroz14} proposed a
more simple configuration including only two
planets, both confirmed using a bayesian analysis of the radial
velocity observations. In the same work they propose the possibility
of a third signal (another planet) with a period of 91 d. We include
this possibility as a true planet.

For our purpose, the three possible configurations are suitable to be
anlyzed dynamically. Our aim is to understand the spin-orbit evolution
of a closely packed planetary system, regardless of the existing number 
of planets. We have a special interest on the GJ 667C system due to the
potential habitability of planet c.

The inclusion of secular perturbations in the
integration of tidal evolution was introduced by Lagrange and Laplace
in the XVIII century. This technique has been widely used to study the
stability of the Solar System (e.g., \citealt{Laskar88}). The
combination of the orbital forcing with tidal effects was also
explored in the specific cases of Venus and Mercury by
\citealt{Correia01, Correia04} respectively. This method was also
applied to exoplanets, first analytically by \cite{Wu02},
\cite{Mardling07}, \cite{Bat09}, \cite{Laskar12}, and after by
performing numerical simulations in \cite{Correia10} and
\cite{Bonfils13}.
  
In order to study the detailed evolution of the orbital parameters, we
have performed numerical simulations of the system
for three possible configurations, spanning a total
time of 50,000 years. In all cases, we will assume that the inclination
of the planetary system, with respect to us, is $\sim 90^o$; hence, 
the observed minimum masses are also the actual planetary masses.
In addition, we are using the barycentric coordinates of the system. 
Figure \ref{fig:SystemSimulation} shows
the evolution of eccentricities in one of our simulations
for the three configurations. These results show
that the dynamic of the system is essentially governed by secular
perturbations, i.e. the semi-major axes and eccentricities remain
varying between a small range of values throughout
all the evolution of the orbits.

However, in the case of 6 planets, 4 of them appear strongly
correlated by pairs, $d$ and $f$, and $c$ and $e$. This is evident in
the ranges of eccentricities and the average value $\bar{e}$ showed in
Table \ref{tab:SystemProperties}.

The 6-planets configuration, is so dynamically packed that orbital
eccentricities oscillate, in some cases, within almost one order of
magnitude (see Table \ref{tab:SystemProperties}).  Planets $b$ and $g$
are the most eccentric, with upper $e_{b}=0.11711$ and $e_{g}=0.11349$
respectively.  Planets like $b$ and $d$ perform the largest
eccentricity excursions, $e_{b} \approx 0.04152-0.11711$ and $e_{d}
\approx 0.00016-0.08145$.

In order to introduce the effect of secular perturbations in the tidal
evolution equations (see Section \ref{sec:TidalEvolution}), we need to
estimate the derivative of the eccentricity $(de/dt)_{\rm pert}$ at
any time for each planetary component.  For that purpose, we first
calculated the Lomb-Scargle periodogram of the numerical
eccentricities for the chosen configurations of the system with: 2, 3 and 6
planets, and identified the periods with the largest power
\citep{Lomb76,Zechmeister09} for the 6-planets configuration. Our
results, in this case, are in good agreement with the analytical
approach made by \citealt{Anglada13}.

Using the principal modes, we reconstructed analytically the eccentricity
and its derivative. Figure \ref{fig:SystemSimulation} also displays 
the evolution of eccentricity and its derivative as reconstructed
analytically with this procedure for the three configurations.

The analysis of the spectral density, realised by means of
Lomb-Scargle's periodograms, indicate that the
fluctuations in the eccentricity are periodic and are dominated only
by a main mode. On the basis of the parameters in table 6 in
\cite{Anglada13}, and using the same symplectic integration package
(HNBody), and equal parameters for the simulation, we performed the
numerical integration of the system to obtain the instantaneous value
of the eccentricity of each planet.  After a Fourier decomposition, we
found an analytical expression that will be used later on the
tidal evolution package.

\section{Tidal Evolution}
\label{sec:TidalEvolution}

Rotational properties of planets and moons (axial tilt and period of
rotation) are affected by their tidal interaction with other bodies.
In general, these interactions produce a net torque $\Tau$ that
modifies the rotational properties and determines their final
equilibrium values. In this work, we assume that the obliquity of the
planets in the system is zero. This assumption is not true, because
the tilt evolution (tilt erosion) is simultaneous to the spin-orbit
evolution. We leave aside this process, but a tilt erosion model is
needed for this system.

We calculated the instantaneous rotational rate $\Omega=\dot{\theta}$, 
if the net torque resulting from the tidal interaction is known, by
solving the equation (see \citealt{Danby62}):

\beq{eq:EOM}
\frac{d\Omega}{dt}=\frac{\Tau}{C},
\eeq

Where $C$ is the moment of inertia around the shortest axis of what
will be in general a triaxial body.  $C$ can be (in general) written
as:

\beq{eq:MoI}
C = \mathcal{C} M_p R^2
\eeq

The net torque on a rotating body experiencing tidal interaction with
a perturber, has in general two contributions: 1) the \textbf{triaxial
  torque} that results when the planet has a triaxial shape and, 2)
the \textbf{tidal torque} arising directly from the strength on the
bulge rose by the perturber.

\subsection{The Triaxial Torque}
\label{subsec:TriaxialTorque}

Triaxiality is mainly due to the inhomogeneous distribution of mass in
a rotating body, as well as the presence of geological features over
the surface.  These induce a triaxial shape in the bodies.

Triaxiality is measured in terms of the three principal moments of
inertia $A,B$ and $C$ where $A<B<C$ and $C$ is in general given by
Eq. (\ref{eq:MoI}).  The triaxial torque depends on the level of
triaxiality as measured by the difference $B-A$ and it is given in
terms of the angle between the sideral direction of the bulge $\theta$
and the true anomaly $\nu$, $\psi=\theta-\nu$. The classic mechanics
expression for this torque could be found in \citet{Danby62} and
\citet{Goldreich66a},

\beq{eq:TriaxialTorque}
\begin{split}
\mathcal{T}_{tri} & = \frac{3}{2} (B-A) \frac{G M}{r^3} \sin 2\psi \\ 
& \approx -\frac{3}{2} (B-A) n^{2}\frac{a^3}{r^3} \sin 2(\theta - \nu),
\end{split}
\eeq

Here $r$ is the instantaneous distance between the centres of the two
bodies.

\subsection{The Tidal Torque}
\label{subsec:TidalEvolution}

In the simplest case of a body with radius $R$, interacting with a
tide-rising perturber of mass $M$ located at an average distance $a$,
the tidal torque is given by the MacDonald (MD) formula
\citep{MacDonald64, Goldreich66, Murray99}:

\beq{eq:MacDonaldTorque} 
\Tau_{\rm tid}=\frac{3}{2}GM^2\frac{R^5}{a^6} k_2\sin
2|\epsilon_g| {\rm Sgn}(\epsilon_g),
\eeq

where $\epsilon_g\equiv(\dot{\nu}-\dot{\theta})\Delta t$ is the
instantaneous geometric lag, i.e. the angle between the direction of
the most elongated part of the planet and the line joining the center
of the bodies (see figure 4 in \citealt{MBE12}) and $\Delta t$ is a
constant time lag.  Here $\nu$ is the orbital true anomaly of the
perturber and $\theta$ is the sideral angle of the planet 
measured with respect to the line of apsides. 

We observe how the tidal torque strongly depends on the
average distance between the bodies $\Tau_{\rm
  tid}\sim\frac{1}{a^{6}}$.  For close planets, such as those found
around GJ 667C, this effect is proven able to erode any initial
rotation period and bring the planet to a low spin-orbit resonance.
The timescale for such spin-down from an initial rotation rate
$\Omega_o$, will be given by the expression from \citet{Gladman96}:

\beq{eq:SpinDownTime}
t_{\rm lock}=\frac{\Omega_o C}{\tau_{\rm tid}}=\frac{2 \Omega_o C a^6 Q}{3 G
  M^2 R^5 k_2},
\eeq

where $Q$ is called the \textit{quality factor}, which is a specific energy
dissipation function. For a rapidly rotating planet $Q\approx1/\sin
2|\epsilon_g|$ \citep{Murray99}.

In a more general case, the tidal torque is obtained from a Fourier
expansion of the tide-rising potential that was first developed by
\citet{Darwin79}, and more recently updated by \citet{Kaula64}.  We
called this formalism the \textit{Darwin-Kaula (DK) expansion}.  The
Fourier modes, over which this expansion is carried out, are called the
tidal modes \citep{EM13},

\beq{eq:modes}
\omega_{lmpq} \approx (l-2p+q)n - m \dot{\theta},
\eeq 

where $lmpq$ are the integers of the Fourier expansion. Here $l$ and
$m$ are the \textit{degree} and \textit{order} of the expansion
respectively, $\dot{\theta}$ is the instantaneous rotation rate of the
body and $n$ is the average orbital angular velocity or the
\textit{mean motion}.  The DK expansion of the potential provides a
rigorous expression for the tidal torque (see equations (109) to (111)
in \citealt{E12a}). The resulting expression has two main components:
an oscillating component that it is averaged out over an orbit, and a
secular component whose average is given by Equation (13) in
\citealt{ME12},

\beq{eq:DarwinKaula} 
\begin{split}
\langle \Ttid \rangle & = 2G M^{2} \sum_{l=2}^{\infty}
\frac{R^{2l+1}}{a^{2l+2}} \sum_{m=0}^{l} \frac{(l-m)!}{(l+m)!}m \ \times \\
 & \sum_{p=0}^{l} F_{lmp}^{2}(i) \sum_{q=-\infty}^{\infty}
G_{lpq}^{2}(e) k_{l}\sin\epsilon_{l}.
\end{split}
\eeq

Here $F_{lmp}(i)$ and $G_{lpq}(e)$ are known as the inclination and
eccentricity functions respectively \citep{Kaula61}.  The most
important feature of the DK formula for the tidal torque, in contrast
to the MD formula in Eq. (\ref{eq:MacDonaldTorque}), is the general
dependence on frequency of the love number $k_l$, the phase lag
$\epsilon_l=\omega_{lmpq}\Delta t_l(\omega_{lmpq})$ and the positive
definite time lag $\Delta t_l(\omega_{lmpq})$.  This dependence on
frequency is key when treating realistically the rheological
properties of planets and moons.

If the bodies are not too close ($R/a<<1$), eccentricity does not
exceed $e\sim0.3$ and we assume a small obliquity $i\approx0$, then we
can truncate the DK torque in Eq. (\ref{eq:DarwinKaula}), leaving only
the term $l=2$, and $q=-7,\cdot,7$.  This gives us an approximate
expression that we will hereafter call the EMW formula,

\beq{eq:EfroimskyTorque}
\begin{split}
\langle\mathcal{T}_{tid}\rangle & = \frac{3}{2}G M^{2}
\frac{R^{5}}{a^{6}} \ \times\\
 & \sum_{q=-7}^{7} G_{20q}^{2}(e)
{\bf k_{2}}(\omega_{q})\sin|\epsilon_{2}(\omega_{q})|\Sgn(\omega_{q}) \ +\\
 & \mathcal{O}(e^8\epsilon) + \mathcal{O}(i^2\epsilon)
\end{split}
\eeq

Here we have written $\omega_{q}=\omega_{220q}$. The Kaula functions
$G_{20q}$ are power series of eccentricity related to Hansen's
coefficients given by \citet{Giacaglia76} via

\beq{eq:Hansen}
G_{20q} = X_{2+q}^{-3,2}.
\eeq  

In order to calculate the tidal torque using the EMW formula, we first 
need to obtain the explicit dependence on frequency of the second
degree Love number $k_2$ and the respective phase lag $\epsilon_2$.  A
rigorous treatment of the problem has been developed in \citet{E12a}
who studied the response of a viscoelastic near-spherical body to
tidal stress.  The result of this careful analysis provide us with the
value of the $k_2$ as a function of planetary bulk properties,

\beq{eq:Love}
k_{2}=\frac{3}{2} \frac{1}{(1+57\mu/8\pi G\rho^2 R^2)},
\eeq 

where $\mu$ is the rigidity and $\rho$ the planetary average density.
On the other hand, the tidal-mode-dependent phase lag $\epsilon_2$ is
given by,

\beq{eq:Keps} 
k_{2}(\omega_{q})\sin \epsilon_{2}(\omega_{q}) =
-\frac{3}{2}\frac{A_{2}
  \mathcal{I}}{(\mathcal{R}+A_{2})^{2}+\mathcal{I}^{2}}\;\Sgn(\omega_{q}),
\eeq

where $\mathcal{R}$ and $\mathcal{I}$ are dimensionless real and
imaginary parts of the complex compliance and $A_{2}$ is a coefficient
which depends on rigidity, radius and mass of the body (see Apenddix B
in \citealt{ME12}):

\beq{eq:R} 
\mathcal{R} = 1 + (\mathcal{X}_q \tau_{A})^{-\alpha}
\ cos\Bigl(\frac{\alpha \pi}{2}\Bigr) \ \Gamma (\alpha +1), 
\eeq

\beq{eq:I} 
\mathcal{I} = -(\mathcal{X}_q \tau_{M})^{-1} - (\mathcal{X}_q
\tau_{A})^{-\alpha} \ sin\Bigl(\frac{\alpha \pi}{2}\Bigr) \ \Gamma
(\alpha +1), 
\eeq

Here $\mathcal{X}_q=|\omega_{q}|\approx|(2+q)n-2\dot{\theta}|$ are the
so called forcing frequencies.  $\alpha$ is a dimensionless parameter
(the Andrade's exponent) that depends on the rheological properties of
the material that makes up the planet.  For materials ranging from ice
to most minerals $\alpha=0.14-0.4$.  $\tau_{A}$ and $\tau_{M}$ are
known as the \textit{inelastic Andrade's time} and the
\textit{viscoelastic Maxwell's time} respectively.  These timescales
characterise the inelastic and viscoelastic properties of the
planetary mantle, and its response to tidal effects in different range
of frequencies (Andrade at high frequencies and Maxwell at low
frequencies respectively).

Our numerical experiments have shown that the assumption for simplicity
$\tau_A=\tau_M$ does not significantly change the results integrating 
the tidal evolution, as well as the possibility of being trapped in a 
given resonance.

\begin{table*}
\centering
\caption{Notes: (1) Values are different for each planet and appear
    in Table \ref{tab:SystemProperties}. (2) Values depend on
    specific numerical experiments.} 
\label{tab:InputParameters}
\begin{tabular}{llllll}
  \hline
	& \textbf{Symbol} & \textbf{Units} &
  \textbf{Earth-like} & \textbf{Waterworld} & \textbf{References} \\ 
  \hline  \hline 

  \multicolumn{6}{l}{\textbf{Planetary Bulk}}\\
  \hline

  Planetary Mass & $\Mp$ & $kg$ & (1) & (1) & \citet{Anglada13} \\
  Planetary Radius & $\Rp$ & $km$ & (1) & (1) & \citet{Zuluaga13} \\
  
  Ice mass fraction  & IMF & -- & 0\% & 50\% & This work \\
  Mantle mass fraction & MMF & -- & 70\% & 20\% & This work \\
  Core mass fraction & CMF & -- & 30\% & 30\% & This work \\
  Triaxiality & $(B-A)/C$ & -- & $5\EXP{-5}$ & $< 10^{-7}$ &  This work \\

  \hline\multicolumn{6}{l}{\textbf{Rheology}}\\
  \hline
  
  Andrade exponent & $\alpha$ & -- & 0.3 & 0.14 & \citet{Efroimsky07}\\
  Bulk Viscosity & $\eta$ & $Pa \ s$ & $10^{20}$ & $10^{18}$ & \citet{Zuluaga13} \\
  Rigidity & $\mu$ & $Pa$ & $8\EXP{10}$ & $8\EXP{9}$ & \citet{Moore03}\\
  Maxwell time & $\tau_{\rm M} \equiv \eta/\mu$ & $yr$ & 40 & 4 & This work \\
  
  \hline\multicolumn{6}{l}{\textbf{Star and Orbit}}\\
  \hline
  
  Mass of the star & $\Ms$ & $M_\odot$ & 0.33 & 0.33 & \citet{Anglada13} \\
  Initial eccentricity & $e_o$ & -- & (1) & (1) &  \citet{Anglada13} \\
  Initial semi-major axis & $a_o$ & $AU$ & (1) & (1) & \citet{Anglada13}\\
  Initial true anomaly/sideral angle & $\theta_o$ & -- & (2) & (2) & \citet{MBE12} \\
  \hline

\end{tabular}
\end{table*}

\subsection{Tidal Energy Dissipation}
\label{subsec:Enegydisp}

Tidal interaction does not only modify the rotation of the planet.  The
work done by tidal forces is dissipated as heat inside the deformed
bodies and, as a consequence, the orbital energy and angular momentum
of the planet change in time. The body acts like a harmonic oscillator 
dissipating energy during each cycle. 

The work done by the drag force, over a displacement during a
given interval of time, results in a rotational energy dissipation during 
each cycle. If the only torque acting on the body is the tidal torque,
the thermal energy released as heat is virtually equal to the work
done by the tide. Therefore, the dissipated energy will be proportional to the phase lag
\citep{FerrazMello08} and tide frequency \citep{E12a}.  Although we
have included a triaxial torque, this force do not produce any net
dissipated energy.  The power dissipated will be simply given by,

\beq{eq:EnergyVariation}
\frac{dE}{dt} = - \sum_{q=-7}^{7} \Tau_{\rm tid,q} \omega_{q} 
\eeq   

Where $\Tau_{\rm tid,q}$ is each of the terms in the sum defining the
average tidal torque in Eq. (\ref{eq:EfroimskyTorque}). The torque
produces a change in the orbital energy of the planet,

\beq{eq:OrbitalEnergy}
E = - \frac{G M_p M}{2a}.
\eeq

and therefore it implies a variation in the semi-major axis given by,

\beq{eq:Semi-MajorVariation}
\frac{da}{dt} = \frac{2a^2}{GM_p M} \frac{dE}{dt}
\eeq

In a similar way the average tidal torque produces a change in the
orbital angular momentum of the planet,

\beq{eq:AngularMomentum}
\mathcal{L} = M_p \sqrt{G M a (1-e^2)},
\eeq

Differentiating and taking into account that $d\mathcal{L}/dt= -
(\langle\Ttid\rangle + \Tau_{\rm tri})$ we found an expression for the
rate of variation of the eccentricity (see eq. \ref{eq:EOM}).

\section{The Model}
\label{sec:Model}

As we have just shown, the rate of rotation and orbital properties of
a planet are modified by the tidal interaction with its host star.
The intensity of this interaction, quantified by the tidal and
triaxial torques, depends on a set of bulk, interior structure and
rheological properties of the planet.  The mass of the central star
and the initial orbital parameters are also required.  In Table
\ref{tab:InputParameters} we summarize the set of input parameters
that are necessary to estimate the net torque on the planets in the GJ
667C system.

Planetary composition and water cycle models 
suggest that planets in GJ 667C could be volatile-rich bodies 
\citep{Schaefer15}. In order to study the 
dependence of the spin-orbit evolution on
composition, we assumed two different sets of input parameters
(columns 4 and 5 in Table \ref{tab:InputParameters}).  The first set
(column 4), hereafter the {\it reference model}, corresponds to planets
with an Earth-like (El) composition, i.e. composed by a silicate
mantle and a metallic core with almost no water content.  The second
set of parameters (column 5) are chosen to match the bulk and
rheological properties of planets containing a significant fraction of
water, i.e.  {\it Waterworlds} (Ww).

Here, we are using the idea of \textit{Ocean planets}, as was proposed
originally by \cite{Kuchner03} and described carefully later on by
\cite{Leager04} and \cite{Adams08}. This description considers planets
with more than 25\% of mass made by water. The internal structure of
this kind of planets is not well known, but the models propose solid
cores made by iron-rock, deep layers made by different water ice
phases, and massive liquid water oceans.

As the bulk structural properties of this Ww are unknown, we consider
that this iron-rocky-ices-water (solid-liquid) planets must have an
internal structure similar to Ganymede \citep{Vance14}.
\cite{Makarov15} called this kind of bodies \textit{semiliquid}, planets
with massive surface oceans and rigid cores that possess a very low
but nonzero triaxiality. In our case, we ran numerical experiments
using a low triaxiality torque for Ww.

To follow the spin-orbit evolution, we need to solve the following set
of coupled, first order, non-linear differential equations: 

\beq{eq:EOM}
\begin{split}
\frac{d\theta}{dt} & = \Omega \\
\frac{d\Omega}{dt} & = [\avTtid(a,e,\Omega)+\Ttri(a,e,\theta,t)]/C \\
\frac{da}{dt} & = -\frac{2a^2}{GM_p M} \sum_{q=-7}^{7} \Tau_{\rm
  tid,q}(a,e,\Omega) \omega_{q} \\
\frac{de}{dt} & = \Bigl(\langle\mathcal{T}_{tid}\rangle + \Tau_{\rm tri}\Bigr)\Bigl(\frac{a(1-e^{2})}{GM}\Bigr)^{1/2}
\frac{1}{eaM_{p}} + \frac{\dot{a}}{2ea} (1-e^{2})
\end{split}
\eeq

It should be noted that these equations only take 
into consideration the tides on the planet. The tides risen by 
the planet on the star are neglected.

With initial conditions $\theta(0)=\theta_o$, $\Omega(0)=\Omega_o$,
$a(0)=a_o$ and $e(0)=e_o$.  The previous set of equations do not
take into account the effect that secular perturbations of other
planets in the system have on the spin-orbit evolution.

\subsection{Numerical Experiments Using TIDEV}
\label{subsec:NumericalExperiments}

In all our numerical experiments on the spin-orbit evolution, we use
a script written in C++ language that we named TIDEV.  The script is
available for public users in \textit{https://github.com/facom/tidev}.

TIDEV is a general framework developed to calculate 
the evolution of rotation for tidally interacting bodies using the 
formalism proposed by EMW. Our numerical tool performs the integration 
of the set of equations (\ref{eq:EOM}). It computes the rotational 
and dynamical evolution of a planet under tidal and triaxial
torques. TIDEV also takes into account the perturbative effects due 
to the presence of the other planets in the system, especially the secular 
variations of the eccentricity.

The input parameters are divided in those related with the bulk 
composition and rheology of the planet and those related with the 
initial orbital configuration of the system.

Typical bulk parameters are the mass and radius of the planet (and 
those of the other planets involved in the integration), the size and 
mass of the host star, the Maxwell time and Andrade's parameter (see Table 
\ref{tab:InputParameters}).

To give the initial orbital configuration, we need to have an initial 
set of orbital elements (see Table 1). The two initial conditions needed 
for each integration are the sidereal angle of the tidal bulge ($\theta$) 
and the initial rate of rotation $\Omega$. This later value is given as a 
function of the mean motion $n$ and depends on the resonance under study; 
for example, if we need to run a simulation in the vicinity of 3:2 resonance, 
the initial rotational velocity reads $\dot{\theta}_0 = 1.51 n$.

For the computation of the locking time, we have 
adopted the assumption that the initial period of rotation is 24 hours. 
Then, the value of $\Omega_0$  for this case is just $\Omega_0 = (2\pi / 86400) \ s^{-1}$.

For our purposes, the most important value returned by TIDEV is the 
time evolution of the rate of rotation of the planet, which enables 
us to estimate the capture probabilities in each resonance considered 
in this work.

As usual in any scheme of numerical integration, determining a 
sufficiently good time step is crucial for the success of the 
integration. With a gross value of the time step (for example, 
if it is comparable with the orbital period), we can lose important 
information about the rotation history of the planet. On the other 
hand, a short time step increases dramatically the computation 
time and also the machine precision becomes a concern. As a general 
rule in our integrations, we need to use a time-step not greater 
than a fraction of a terrestrial day. In spite of this, even with 
this selection of the time step, the integrations are expensive 
in a computational sense.

A new numerical tool was developed 
and presented by \citet{Bolmont15}. Their model is based on the classic 
programme Mercury for N-bodies, and has been extended to include the 
effects of the tidal torque, the deforming effect of the rotation and 
the relativistic effects produced by the nearness of the planet to its 
star.

The tidal model used by \citet{Bolmont15} is the classic 
model of constant lag by \cite{Hut81}. Though the model does 
not define explicitly the tidal torque, it is expressed in terms 
of momentum (see equation (5) in \cite{Bolmont15}). Unlike our model, 
theirs lack of the possible effects that would 
have a change in the bulk composition of the planets, and therefore, 
their model does not include changes in the rheology. 
Also, it does not include calculations of the possible effects caused by 
other planetary members in the system, this is, additional gravitational 
disturbances.

On the other hand, their model calculates, like ours, 
the time scale that a planet takes to be tidally locked as well 
as the periods of rotation reached at the end of the spin-orbit evolution.

\begin{figure*}  
  \centering
   \subfigure{\includegraphics[width=85mm]{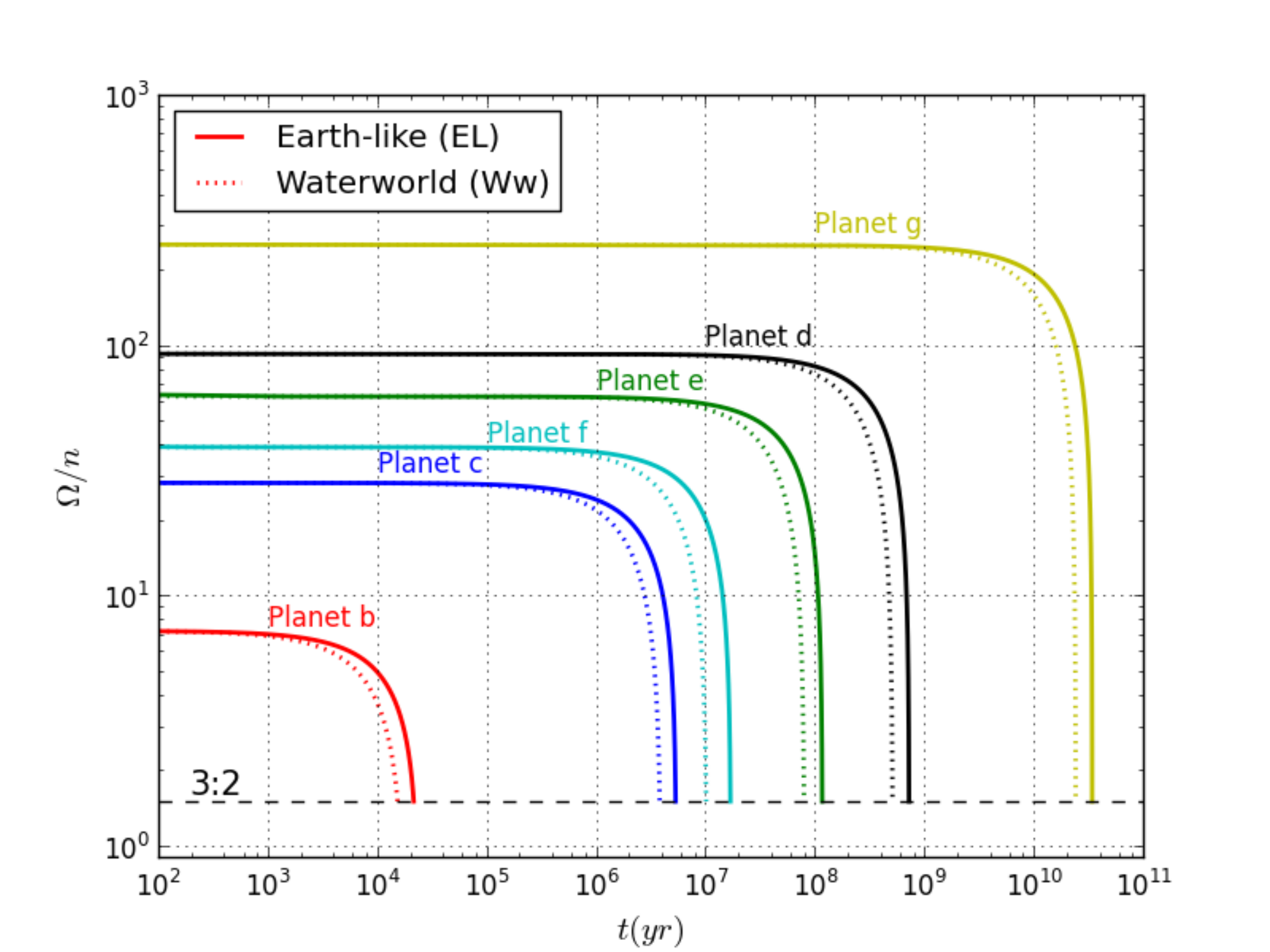}}
   \subfigure{\includegraphics[width=85mm]{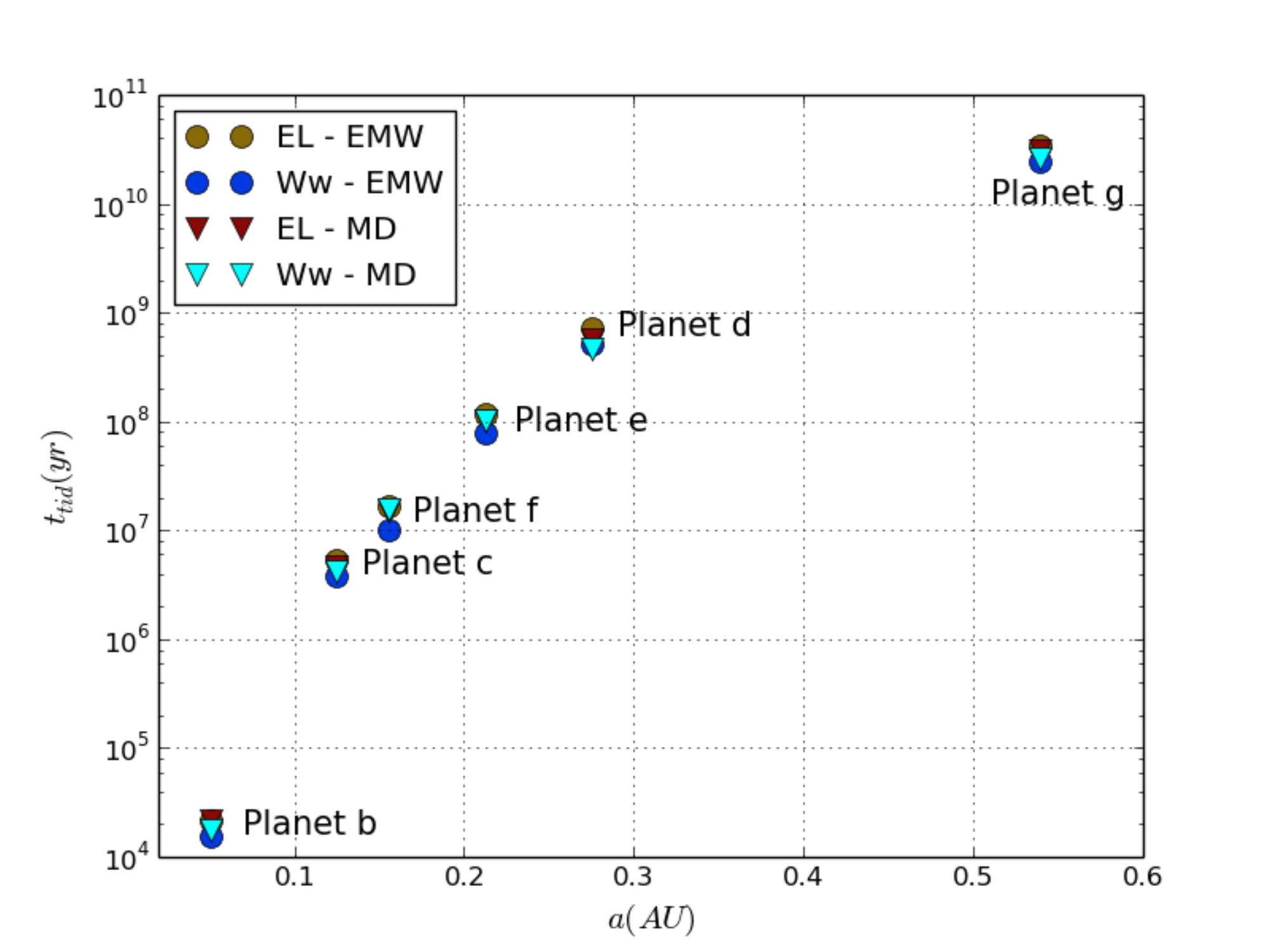}}
   \caption{Left panel: spin down times for El (continuos
     line) and Ww (doted line) planets.  Right panel:
     Spin-down time comparison between the numerical integration using 
     EMW and MD models.}
   \label{fig:EquilibriumEccentricity}
\end{figure*}

The main disadvantage in the work of \cite{Bolmont15} is 
the frequency-independent tidal model. Their approach 
introduce the rotational evolution separated from the tidal forces, 
the deformation and the relativistic effect, which is not ideal 
from our point of view. In our case, the calculation of gravitational 
interactions between planets was done previously, then 
we included in the rotational evolution integration.

In order to make a simple quantitative comparisson 
between Bolmont's model and our model, we made the same computations 
for Kepler-62 b with the parameters given in table (7) of \citet{Bolmont15} 
for the semimajor axis and eccentricity. Regarding the 
bulk properties of the planet, we assume a mass 5\% higher
than the nominal value as explained in their work (i.e. 
equal to 2.72 $M_{\oplus}$) and a radius of 1.31 $R_{\oplus}$. 
The rheological parameters were taken equal to those of an 
El planet, listed in Table \ref{tab:InputParameters} of this work. 
We computed the capture probabilities around 3:2 and 1:1 resonances and
obtained a capture probability of 41.45\% in 3:2 resonances yielding
a 58.55\% probability of being captured in the synchronous 
resonance. These results support the idea (only for planet Kepler-62 b) 
presented in Bolmont's work that the three inner planets of the 
systems are slow rotators with rotation periods grater than 
100 hrs. Similarly, our results for the locking time are in the same 
order of that from Bolmont ($10^3 \ yr$). We need to clarify that in 
these simulations we did not take into account the evolution of the 
obliquity, only that of the rotational period.
 
Turning to our model, there are two interesting quantities to 
compute for each planet: (1) the {\bf spin-down time, $t_{lock}$}, 
i.e. the time required for an
almost complete erosion of any initial rotation rate $\Omega_o$ until it
reaches a resonant state where the rotational angular velocity is close
to the orbital angular velocity; (2) the {\bf capture probability
  $P_{c}$}, i.e. the fraction of times the planet is trapped in a
given spin-orbit isolated resonance where rotation and orbital angular
velocities become commensurable, $\Omega/n\approx(2+q)/2$ with
$q=0,1,2\cdot$. As a side note, the $1/2$ spin-orbit 
resonance ($q=-1$) is also possible under certain conditions, 
if the planet evolves from an initial retrograde spin.

We calculate the spin-down times integrating the set of Equations
(\ref{eq:EOM}) in a coarse grained time grid.  We have verified that
semi-major axis and eccentricities are not significantly modified by
tidal interaction in this particular system.  Therefore, only the
first two equations should be solved.  If we assume a relatively large
value for $\Omega_{o}$, i.e. $\Omega_{o}=7.3 \times 10^{-5} \ s^{-1}$, 
the tidal and triaxial torques are small and hence, we
can assume, for a given range of time, an average constant tidal torque.

Table \ref{tab:SDR} shows the spin-down time for a
6 planets configuration and its dependence on the
set of planetary properties we assumed (El or Ww planets). For each
planet we have also calculated the $t_{lock}$ using the classical
MacDonald's parameters (see Equation \ref{eq:SpinDownTime}).

\begin{table}
\centering
\caption{Spin-down times in years for all planets in the GJ 667C system.
  Times were calculated using both, EMW and MD methods. For EMW 
  calculations, we computed the value of the Love
  number $k_2$ using eq. (\ref{eq:Love}) and calculated an effective
  $Q_{e}$ for each $t_{lock}$. For MD calculations, we used $k_2$ and
  $Q$ values from literature: \citealt{Henning09} for Earth-like planets and 
  \citealt{Barnes12} for Waterworlds.} 
\label{tab:SDR} 
 
\begin{tabular}{cccc}
 \hline 
                  & $t_{lock}$ (yr)  & $k_2$ & $Q_{e}$ \\
 \hline  \hline 
 Earth-like - EMW & & & \\
 \hline
 \textbf{b}       & $2.16 \times 10^4$ &  1.05 & 173.55 \\
		  
 \textbf{c}       & $5.34 \times 10^6$ &  0.92 & 174.47 \\
		  
 \textbf{f}       & $1.69 \times 10^7$ &  0.68 & 125.47 \\
		  		  
 \textbf{e}       & $1.16 \times 10^8$ &  0.79 & 146.25 \\
		  
 \textbf{d}       & $7.25 \times 10^8$ &  1.01 & 214.90 \\
		  
 \textbf{g}       & $3.42 \times 10^{10}$ &  0.96 & 178.33 \\
 \hline
 Earth-like - MD  & & & \\
 \hline
 \textbf{b}       & $2.17 \times 10^4$ &  0.3 & 50 \\
		  
 \textbf{c}       & $4.67 \times 10^6$ &  0.3 & 50 \\
		  
 \textbf{f}       & $1.53 \times 10^7$ &  0.3 & 50 \\
		  		  
 \textbf{e}       & $1.04 \times 10^8$ &  0.3 & 50 \\
		  
 \textbf{d}       & $5.68 \times 10^8$ &  0.3 & 50 \\
		  
 \textbf{g}       & $3.06 \times 10^{10}$ &  0.3 & 50 \\
 \hline
 Waterworld - EMW& & & \\
 \hline		 
 \textbf{b}       & $1.53 \times 10^4$ & 1.39 & 343.55 \\

 \textbf{c}       & $3.82 \times 10^6$ & 1.34 & 243.35 \\ 

 \textbf{f}       & $1.01 \times 10^7$ & 1.31 & 166.67 \\

 \textbf{e}       & $7.96 \times 10^7$ & 1.34 & 211.06 \\

 \textbf{d}       & $5.12 \times 10^8$ & 1.39 & 301.23 \\

 \textbf{g}       & $2.42 \times 10^{10}$ & 1.37 & 251.96 \\
 \hline
 Waterworld - MD & & & \\
 \hline		  
 \textbf{b}       & $1.75 \times 10^4$ & 0.5 & 100 \\

 \textbf{c}       & $4.15 \times 10^6$ & 0.5 & 100 \\

 \textbf{f}       & $1.58 \times 10^7$ & 0.5 & 100 \\

 \textbf{e}       & $1.01 \times 10^8$ & 0.5 & 100 \\

 \textbf{d}       & $4.72 \times 10^8$ & 0.5 & 100 \\

 \textbf{g}       & $2.64 \times 10^{10}$ & 0.5 & 100 \\
 \hline 
\end{tabular}
\end{table}

\begin{figure*}  
   \centering
    \subfigure{\includegraphics[width=85mm]{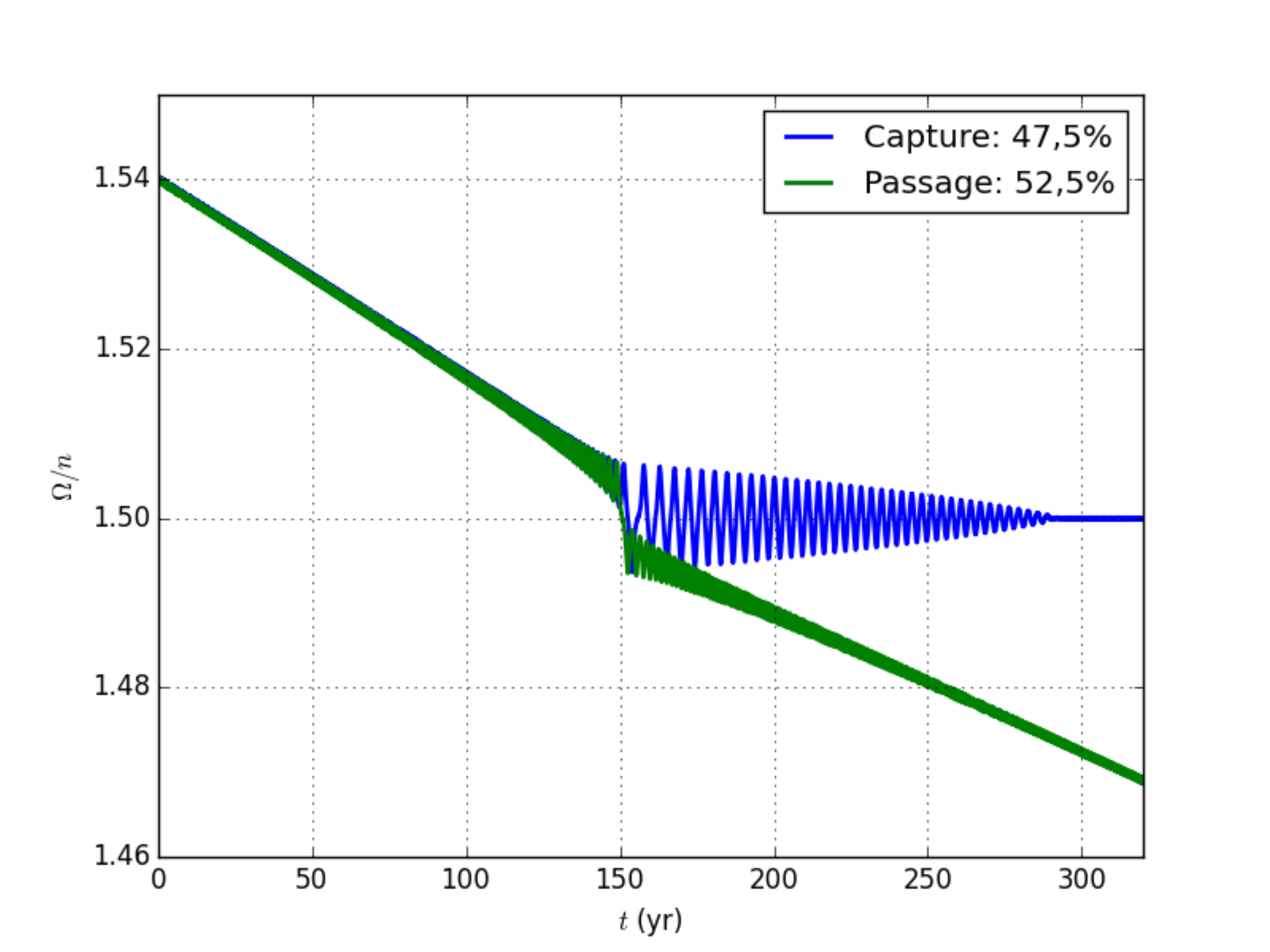}}
    \subfigure{\includegraphics[width=85mm]{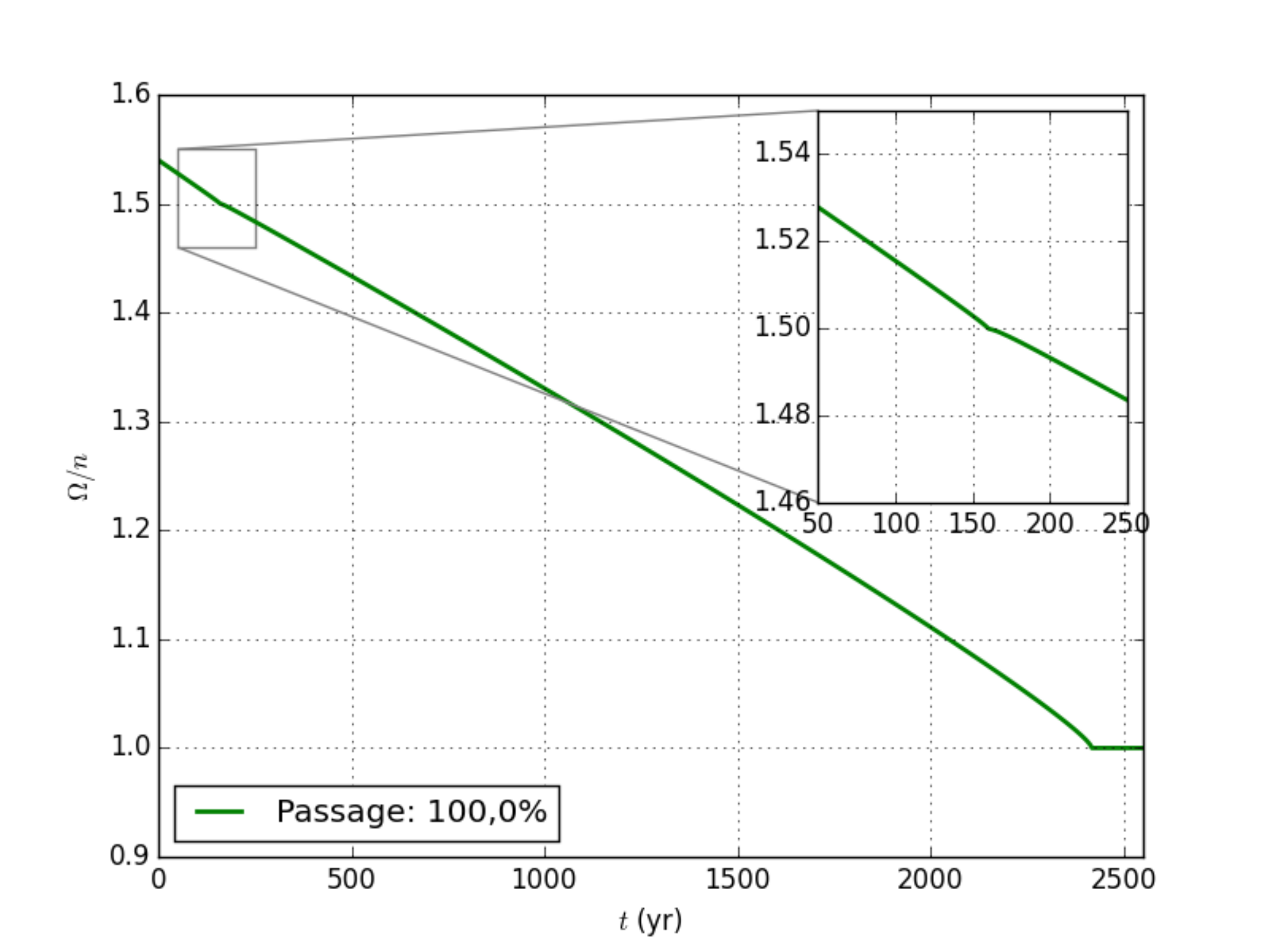}}
   \caption{Probabilities to reach 3:2 resonance for Planet $b$, Earth-like
      (left panel) and Waterworld (right panel).}
   \label{fig:probabilities}
\end{figure*}

\section{Results}
\label{sec:Results}

Our main results are related with the probability to reach low
resonances and its dependence on composition, changes on eccentricity
and interactions with the other bodies in the system.

First, we analyse the influence of the secular changes in the orbital
elements, i.e. changes in eccentricity and semi-major axes (see
Section \ref{sec:GJ 667C}).  On the other hand, we developed numerical
experiments to determine the dependence of the model on fictitious
eccentricity changes.
    
\subsection{Capture Probabilities}
\label{subsec:CapProb}

It is known that the probability of capture depends on the
eccentricity and other quantities as the quality factor. We have
calculated the capture probability of different spin-orbit
resonances. For that purpose we use the \textit{brute force} method used
previously by \citet{Correia04} and \citet{MBE12}, that for the sake
of completeness we will explain here.  We start with an initial
condition $\Omega(0)$ close to the studied isolated resonance, i.e.
$\Omega(0)/n = (2+\eta)/2+\epsilon$.

Around the isolated resonance, the tidal torque becomes minimum and, as
a consequence, the system tends to preserve this idealised state.
However, as the planet naturally crosses the resonance, its angular
velocity starts to oscillate around the resonant frequency
(librations).  According to the initial value assumed for $\theta$,
the sideral angle formed between the bulge axis and the star, the
chances the planet has to traverse this resonance and emerge without
being trapped is different.

This is exactly what makes this process stochastic in nature.  The
number of times the planet is trapped in a given resonance defines the
probability of that resonance. We did 40 runs for each initial
sidereal angle, for each planet and for two different compositions
each time. The same was made to include the interactions with
other planets for two different configurations of the system: 3 and 6
planets.

\begin{figure*}  
   \centering
   \subfigure{\includegraphics[width=85mm]{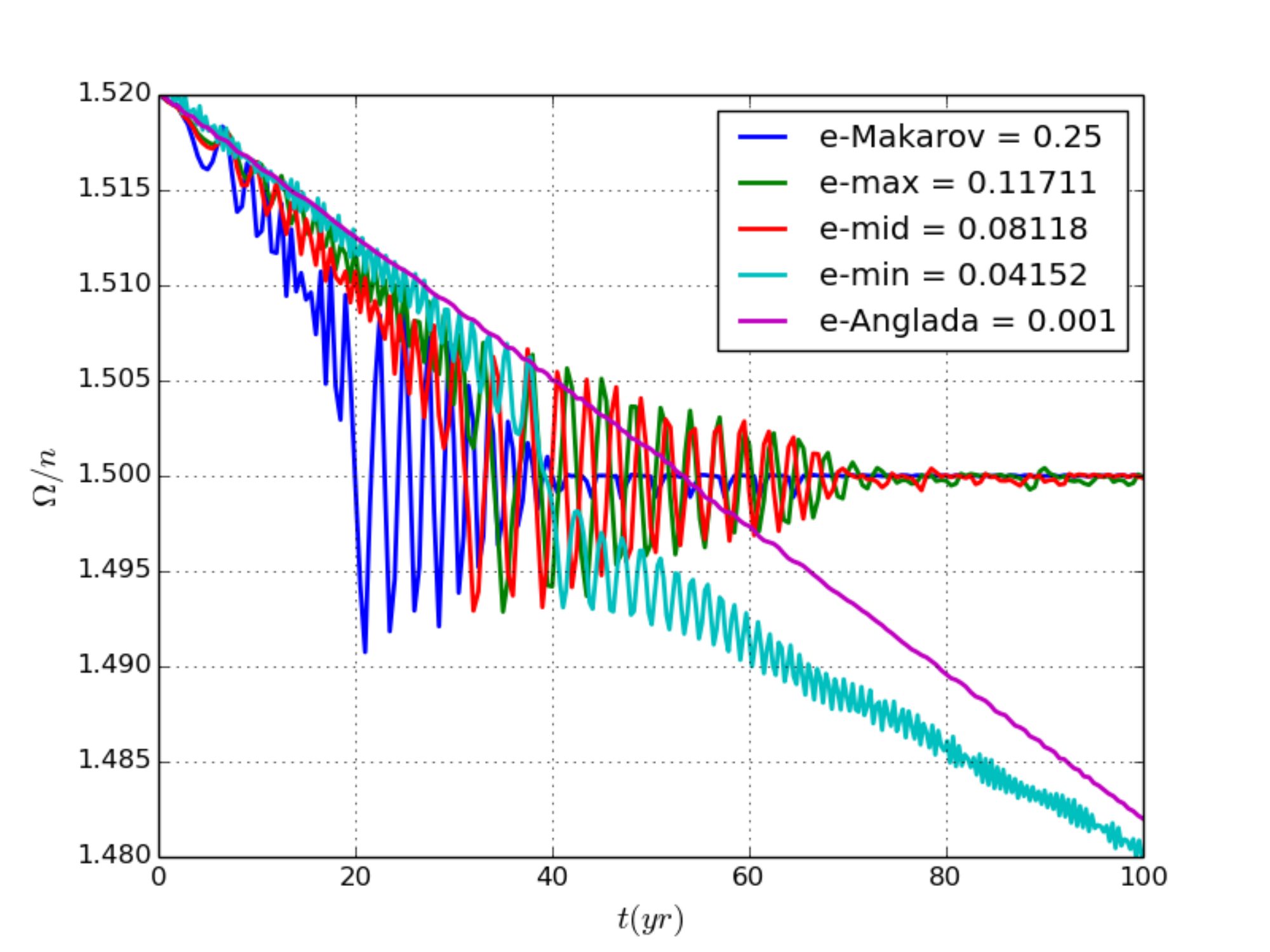}}
   \subfigure{\includegraphics[width=85mm]{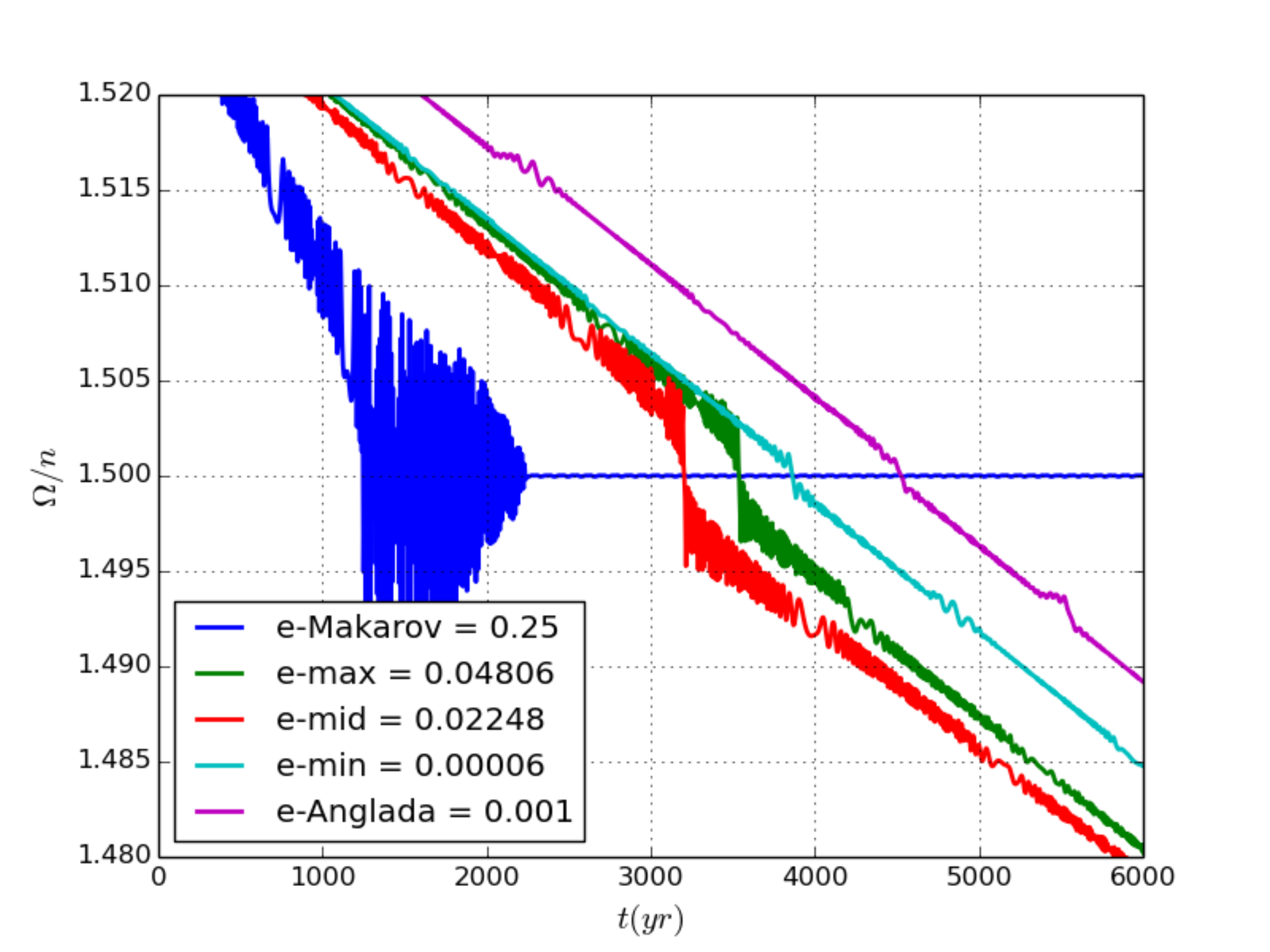}}
   \caption{Earth-like planets $b$ (left) and $c$ (right) arriving and passing 3:2
     resonance for different eccentricities.}
   \label{fig:arriving}
\end{figure*}

\begin{table}
\centering
\caption{Probability of capture for planets b and c, for Earth-like (El) and 
Waterworld (Ww) composition for three different configurations: 1) Star-planet, 
2) Star-planet + 2 more planets (including perturbations), 
and 3) Star-planet + 5 more planets. The eccentricity used for El 
planets is \textbf{$\bar{e}$} (see Table 1).}   
\label{tab:ResonanceCapture}
  \begin{tabular}{lcccc}
  \hline  
 \hline
 & e & 3:2 & 2:1 & 5:2\\
 \hline
  & & \textbf{Star - planet}& & \\
 \hline \hline 
 \textbf{El b} & 0,08 & 47,5\% & 10,2\% &  0,0\% \\
 \textbf{El c} & 0,02 & 10,0\% & 0,0\%  &  0,0\% \\
 \hline
 \textbf{Ww b} & 0,08 & 0,0\% & 0,0\% &  0,0\% \\
 \textbf{Ww c} & 0,02 & 0,0\%  & 0,0\%  &  0,0\% \\
 \hline
  & & \textbf{Star - 3 planets}& & \\
 \hline \hline
 \textbf{El b} & 0,08 & 41,0\%  & 15,0\% &  0,0\% \\
 \textbf{El c} & 0,02 & 0,0\%   & 0,0\%  &  0,0\% \\
 \hline
 & & \textbf{Star - 6 planets}& & \\
 \hline \hline
 \textbf{El b} & 0,08 & 41,0\%  & 15,4\% &  0,0\% \\
 \textbf{El c} & 0,02 & 0,0\%   & 0,0\%  &  0,0\% \\ 
  \hline \hline
  \end{tabular}
  \label{tab:probabilidad}  
\end{table}

In Table \ref{tab:ResonanceCapture}, we show the results of the
calculated probabilities of capture in three isolated resonances (5:2,
2:1, 3:2), for planets $b$ and $c$. The table shows the results of 
including the tidal interactions between the planet and the host star 
and the gravitational interaction of that planet with other bodies in 
the system, in this case, including a configuration of 3 and 6 planets.

For planet $b$ (the closest planet), the dependence of resonance
capture on bulk composition is clear, when you change from El to Ww
composition, the probability of capture down to zero for any resonance
higher than 1:1.  For an El composition, this planet reaches the 3:2
resonance in 47.5\% of the cases in a period of time less than 22 Kyr.

When other planets are included, the probability varies from 47.5\%
to only 41\% for resonance 3:2, but for resonance 2:1 the probability
increases from 10.2\% to 15.0\%. These changes are caused by the
periodic variations in eccentricity when other bodies are added to the
system.

In the case of planet $c$, for El composition the planet ends 
10.0\% of the cases in a resonance 3:2, in a time less than 5.5
million years.

For the Ww case, and for both planets, $b$ and $c$, the probability of
capture in any resonance is 0.0\% for low eccentricities.

Considering that the actual composition of the planets remains unknown, 
these numerical experiments just address us to an idea about the behaviour of
the real planets. However, the rheology of the
planet provides us with information about the tendency of planets with
different viscoelastic behaviour to becoming trapped in a specific
resonance.  At first sight, planets with low average density tend to
cross resonances like 2:1 or 3:2 and get the 1:1 resonance directly.

On the other hand, although the triaxiality of the body produces a measurable torque,
after analysing the dynamic evolution of the system and obtaining the
calculations of probability for a 3:2 resonance with the EL planet
$c$, we can assure that the contribution of the triaxial torque to the 
secular evolution of eccentricity, is negligible. In Figure \ref{fig:triax} we can see 
planet $c$ arriving to the resonance 3:2 in both cases, with and
without considering the triaxial torque. The probabilities 
of capture are not modified if we includes the triaxial torque. 

In the case of Waterworlds, we developed numerical experiments within
a range of low triaxialities ($< 10^{-7}$), even zero in some
cases, and we found that for this kind of planets the probability 
of capture in any resonace higher than 1:1 is zero for low 
eccentricities (see table \ref{tab:probabilidad}).

When you have low values of the triaxiality, capture in a given
resonance is achieved for smaller values of the eccentricity (see fig
8 in \citealt{M12}).  In other words, some resonance of low order is
more likely to occur if the traxiality tends to zero, which has been
verified with our calculations. 


\begin{figure}  
   \centering
   \includegraphics[width=85mm]{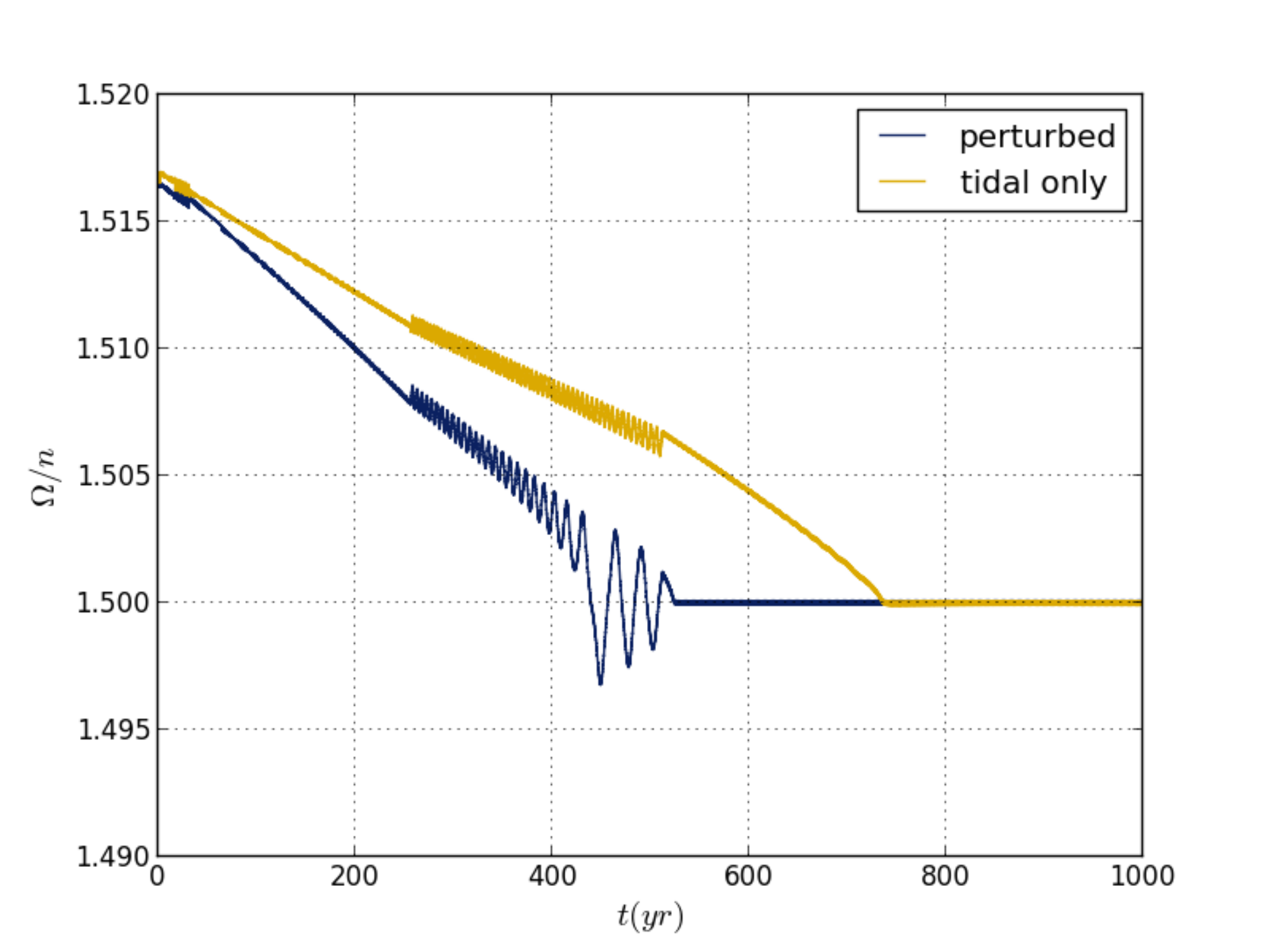}
   \caption{El planet $c$ perturbed gravitationally by all the other members 
     of the system (6 planets configuration) 
     compared with a non perturbed (tidal only) Star-planet configuration.}
   \label{fig:perturbed}
\end{figure}

\subsection{Changes on the Eccentricity}
\label{subsec:changesEcc}

The effects of the eccentricity variation produced by the
gravitational interaction between the planets have been introduced in
the rotational evolution, entered through an analytical expression
obtained from a Fourier transformation of the eccentricities, obtained
in turn through a numerical integration (see Section \ref{sec:GJ
  667C}). It is important to note that the time in which the
eccentricities evolve is much larger than the orbital and rotation
rates of the planets.

Therefore, to evaluate the impact of the eccentricity on the capture
probabilities, first we took three representative values of
eccentricity for each planet: $e_{min}, \bar{e}, e_{max}$ (see Table
\ref{tab:SystemProperties}), and studied the evolution of the
rotational rate of various spin-orbit resonances. We found that in any
case the final captured probabilities were not altered. We include
calculations made with other values of the eccentricity obtained by
\citet{Anglada13} and \citet{MB14} also (see Figure
\ref{fig:arriving}).

A particularly interesting dynamical scenario arises when mean tidal
torque vanishes. This situation leads to a state of dynamical
equilibrium where the disturbed body spins with constant angular
velocity \citep{M12}. 

Depending on the form of the torque, we can
infer the values of eccentricity for which it vanishes giving a fixed
value of $\dot \theta$. The first model is the Constant Time
Lag (MD model) which sets the values of the time delay $\Delta t$ as
independent of the tidal mode frequency. In this model, the mean tidal
torque involves the following terms \citep{Hut81,ME12}:

\begin{equation}
\label{eq001}
\displaystyle <\tau_{tid}> \propto \left[ \frac{1+\frac{5}{2}e^2 +
    \frac{45}{8}e^4 + \frac{5}{16}e^6 }{(1-e^2)^6} -
  \frac{\dot{\theta}}{n} \frac{1+3e^2 + \frac{3}{8}e^4}{(1-e^2)^{9/2}}
  \right]
\end{equation}

\begin{figure}  
  \centering
  \includegraphics[width=85mm]{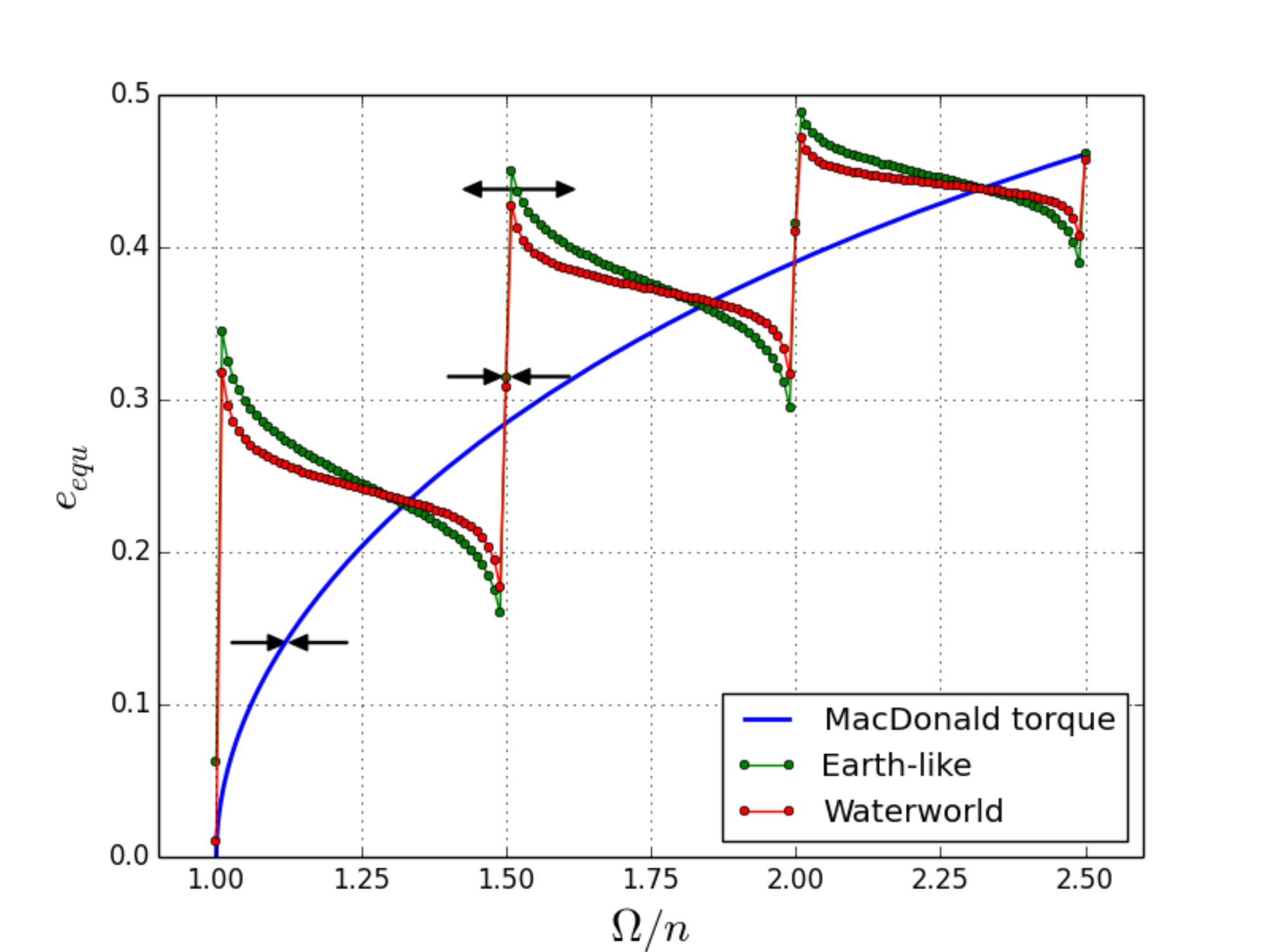}
  \caption{Resonance depending on an equilibrium eccentricity for both 
  compositions in the case of planet $c$. }
  \label{fig:Pvse}
\end{figure}

\begin{figure}  
   \centering
   \includegraphics[width=85mm]{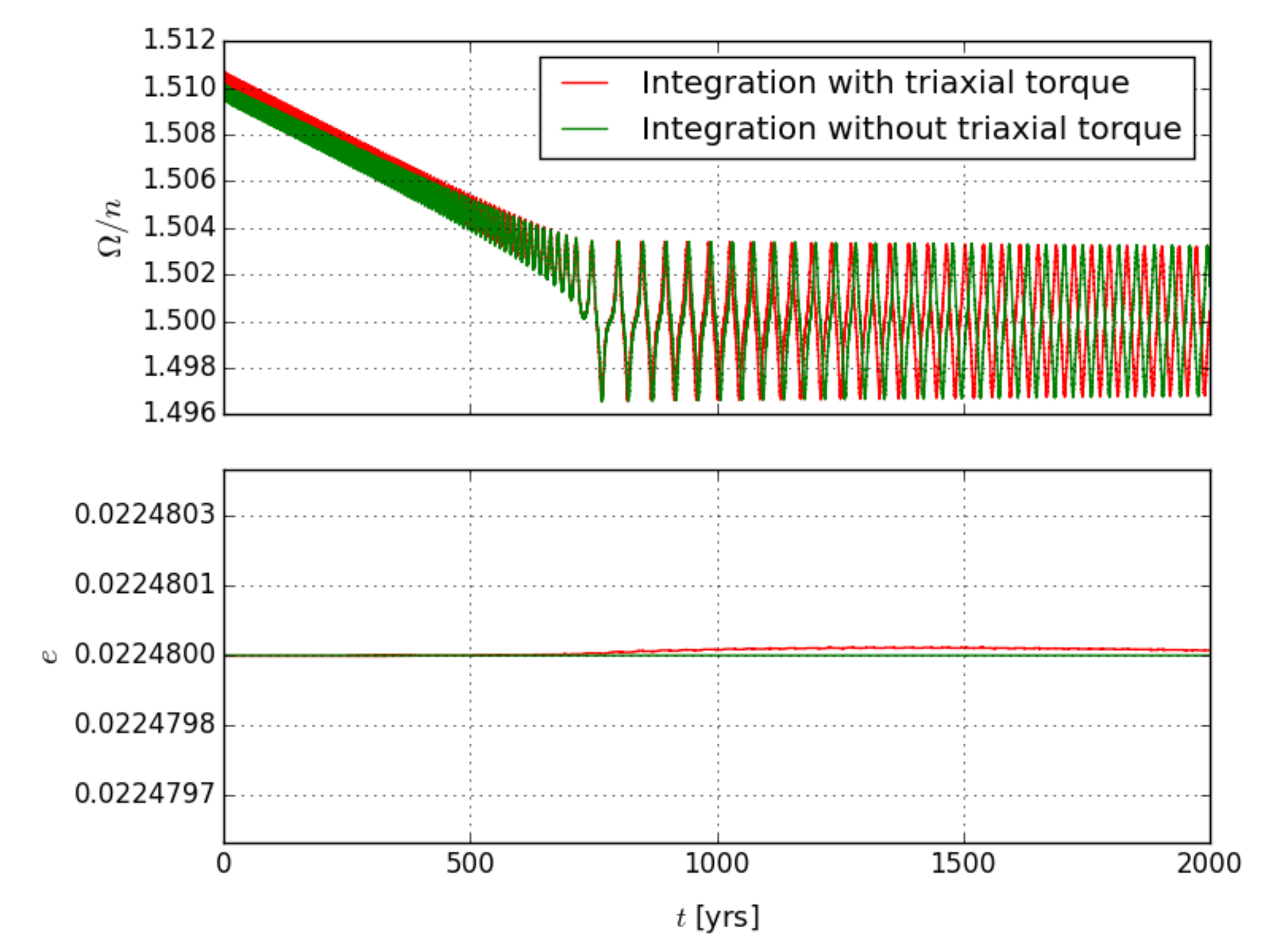}
   \caption{Earth-like planet $c$ eccentricity evolution, including 
   both, tidal and triaxial torques.}
   \label{fig:triax}
\end{figure}

By setting $<\tau_{tid}> = 0$ we obtained the following approximate
expression for the angular velocity as a function of the eccentricity
and the mean motion taken as a constant

\begin{equation}
\label{eq002}
\displaystyle 	\dot{\theta} = n \left[ 1+6e^2 + \frac{3}{8}e^4 + \frac{173}{8}e^6 + \mathcal{O}(e^8)\right ]
\end{equation}

which can be solved for the eccentricity for a given resonant or
non-resonant value of $\dot {\theta}/n$. This root is called the
equilibrium eccentricity. The behaviour of this eccentricity is
depicted by the blue rising curve in Figure \ref{fig:Pvse}. The region
above the blue line corresponds to positive values of the mean torque;
therefore, in this zone the torque accelerates the rotational motion of
the body. Below the blue line, the torque is negative and it
contributes to despin the body. In this figure we can see the 
intrinsically stable behaviour of the equilibrium states in the sense that any
horizontal perturbation made on a point of the curve (increasing or
decreasing the rotational velocity) is counteracted by a torque which
tends to return the motion to the original state as depicted by the
direction of the black arrows.

We proceeded in the same way with the EMW form of the tidal torque
in order to explore the concept of equilibrium eccentricity in this
model. We solved Equation (\ref{eq:EfroimskyTorque}) for the
eccentricity contained inside the $G_{20q}$ functions for a set of
resonant and non-resonant values of $\dot{\theta}/n$ ranging from 1:1
to 5:2.  As we can see from the red and green curves in Figure
\ref{fig:Pvse}, the behaviour of equilibrium eccentricity is completely
different in this case. Perturbing any arbitrary but non-resonant
point of the curve, leads to torques which act in the same direction of
the perturbation in contrast to the MD case. Therefore, the
equilibrium of points outside any resonance is unstable by nature in
this formalism. On the other hand, when we disturb a resonant point
(the central point along the nearly vertical lines), two torques emerge
in opposite direction to the perturbation counteracting its effect.
From this discussion, in the MD model any possible $\dot\theta/n$
represents a state of stable equilibrium, but with the EMW torque,
only resonant values are stable, if are coupled with the appropriate 
value of eccentricity. In both, the MD and in the EMW
torque, we only have one eccentricity value for which the average
torque is zero, but this does not mean that it is stable in the last
case (see Section 3 in \cite{ME12}).

In an empirical sense, we can make general conclusions about the final
spin state of a body from the equilibrium eccentricity in Figure
\ref{fig:Pvse}. In fact, if we introduce a fictitious perturbation
that covers a range of eccentricities wide enough to contain one of
the resonant and stable values of the angular velocity, the
probability of capture in that resonance is no longer zero no matter
how high the resonance is. We made an experiment with planet $c$
introducing a fictitious sinusoidal perturbation with an amplitude of
0.4. We found that the most probable resonance is 3:2 with a
probability of 85\%. But interestingly, the next most probable
resonance was 2:1 with the remaining 15\% of the probability. This
experiment confirms our suspicion that providing a wide perturbation
to the eccentricity, there could be capture in high order 
resonances. This idea was discussed previously in classic
works as \citet{Goldreich66a}.

\subsection{Semi-analytical Probability}
\label{subsec:semianal}

Another semi-analytical method to get a probability of capture, is
considering the librations around the resonance as proposed by
\citet{Goldreich66} and \citet{Goldreich68}. In this method, the
capture depends on the change of the kinetic energy at the end of the
libration and its relation with the total energy dissipation in a
complete cycle.

The semi-analytical formula for the probability derived by
\citet{Goldreich66} adapted to the DK torque gives:

\begin{equation}
\label{eq01}
\displaystyle P_{capt} = \frac{2}{ 1+2\pi V/\int_{-\pi}^{\pi}W(\dot{\gamma})d\gamma }
\end{equation}

where 

\begin{equation}
\label{eq02}
\displaystyle V = K\sum_{q\neq q'}G_{220q}^2k_2((q-q')n)\sin |\epsilon_2((q-q')n)| \text{Sgn}(q-q')
\end{equation}

and

\begin{equation}
\label{eq03}
\displaystyle W(\dot{\gamma}) = -KG_{220q'}^2 k_2(\dot{\gamma}) \sin |\epsilon_2(\dot{\gamma})|\text{Sgn}(\dot{\gamma}) 
\end{equation}

being $K$ a positive constant which does not play a role in the
computation of capture probability.  Numerical integration of this
equation for different values of eccentricity gives us insight on the
strong dependence of both quantities. In fact, for planet $b$, we 
found that the capture in 3:2 resonance is guaranteed for eccentricities 
equal or greater than 0.15. On the other hand, the probability of capture 
in 2:1 is almost 23\% for that same eccentricity ($e=0.15$), then the capture in this
resonance is also possible (see Figure \ref{fig:eccvsres}).

To evaluate this integral, we must consider the following separatrix
equation which relates the parameters $\dot \gamma$ and $\gamma$, which
define the phase space trajectory near a given resonance.

\begin{equation}
\label{eq04}
\displaystyle \dot{\gamma} = 2n \left[ 3\frac{(B-A)}{C} G_{20q'}(e) \right]^{1/2} \cos \frac{\gamma}{2}
\end{equation}

We performed a series of experiments with El planets $b$ and $c$ to
compare both methods: the brute force and the semi-analytical one.
For planet $b$, with the former method, we obtained a capture
probability of 47.5\% in the 3:2 resonance and, with the
semi-analytical method, we obtained for the same planet and resonance
a probability of $\sim 50.0\%$, which represents a reasonably good
agreement between both methods (see Figure \ref{fig:eccvsres}). The
same comparison was performed with planet $c$ and we got 10.0\% and
$\sim 6.0\%$ of capture probability in 3:2 resonance with the two
methods respectively.

This semi-analytical method was also applied to the Ww versions of
planets $b$ and $c$, but the results in this case were quite different
from those obtained by brute force. However, we do not believe the
semi-analytical formula can be used in this case because it depends
strongly on the behaviour of $(B-A)/C$. In fact, the separatrix
equation only takes into account the triaxial torque, but not the tidal
torque. The effects of the last one are estimated by the energy
dissipation calculated using the phase space trajectory which is
determined by $(B-A)/C$. Since Ww planets do not have a permanent
triaxiality, equations (\ref{eq01}) and (\ref{eq04}) are inappropriate
for computation of the isolated resonance capture probability
(Makarov, personal communication). Then, the comparison between two
methods carried out in this work is limited to El planets.
  
\begin{figure}  
\centering
\includegraphics[width=85mm]{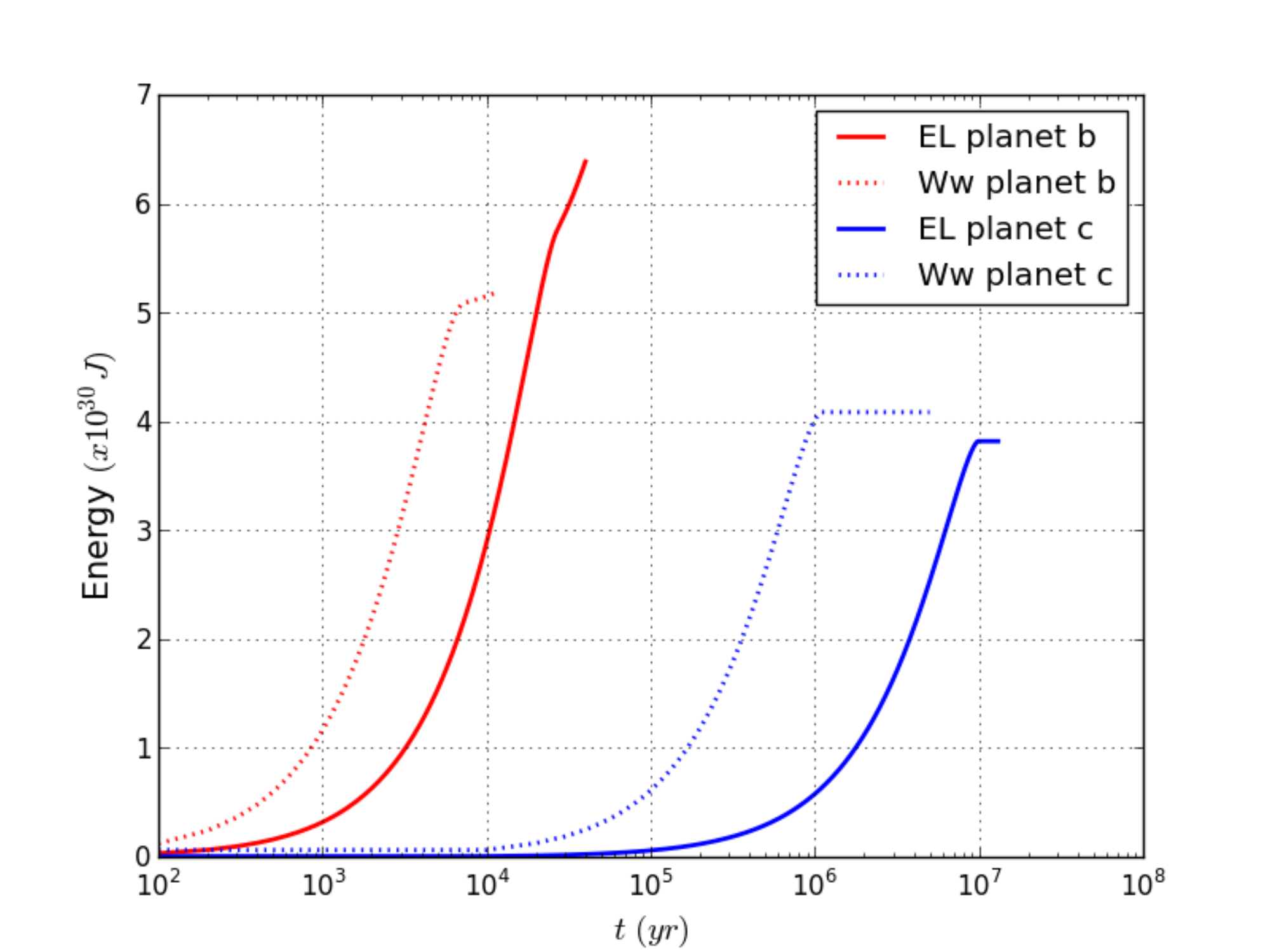}
\caption{Tidal energy dissipated by El and Ww planets
  $b$ and $c$.}
\label{fig:energy}
\end{figure}

\subsection{Energy Dissipation}
\label{subsec:Energy}

We used the results for the tidal energy dissipation to compare the
times of locking for the two different compositions. In Figure
\ref{fig:energy} we show how El planets take more time to dissipate
the tidal energy and finally reach the resonance.

But according to the classical expression (see
eq. (\ref{eq:SpinDownTime})), this time is directly proportional to
the quality factor $Q$. Why do Ww, that have a greater $Q$, take less
time to reach the resonance? We deduced that in this case the size of
the planet has more effect over the spin-down time than the quality
factor.

As we can see in Table \ref{tab:SDR}, the effective quality
factor for Ww are 2 times greater than the El planets. The spin-down
times for these planets should also be 2 times greater. On the other
hand, $t_{lock} \sim R^{-5}$, according to this, bigger planets
take less time to reach a final low resonance. They dissipate tidal
energy rapidly compared to small planets, even when having a greater $Q$.

As a result, the dissipation of tidal energy is related to the
planetary composition and its rheology. According to our results,
water-rich planets achieve the low resonance or the total tidally
locked state, far less than rocky planets.

\begin{figure}  
  \centering
  \includegraphics[width=85mm]{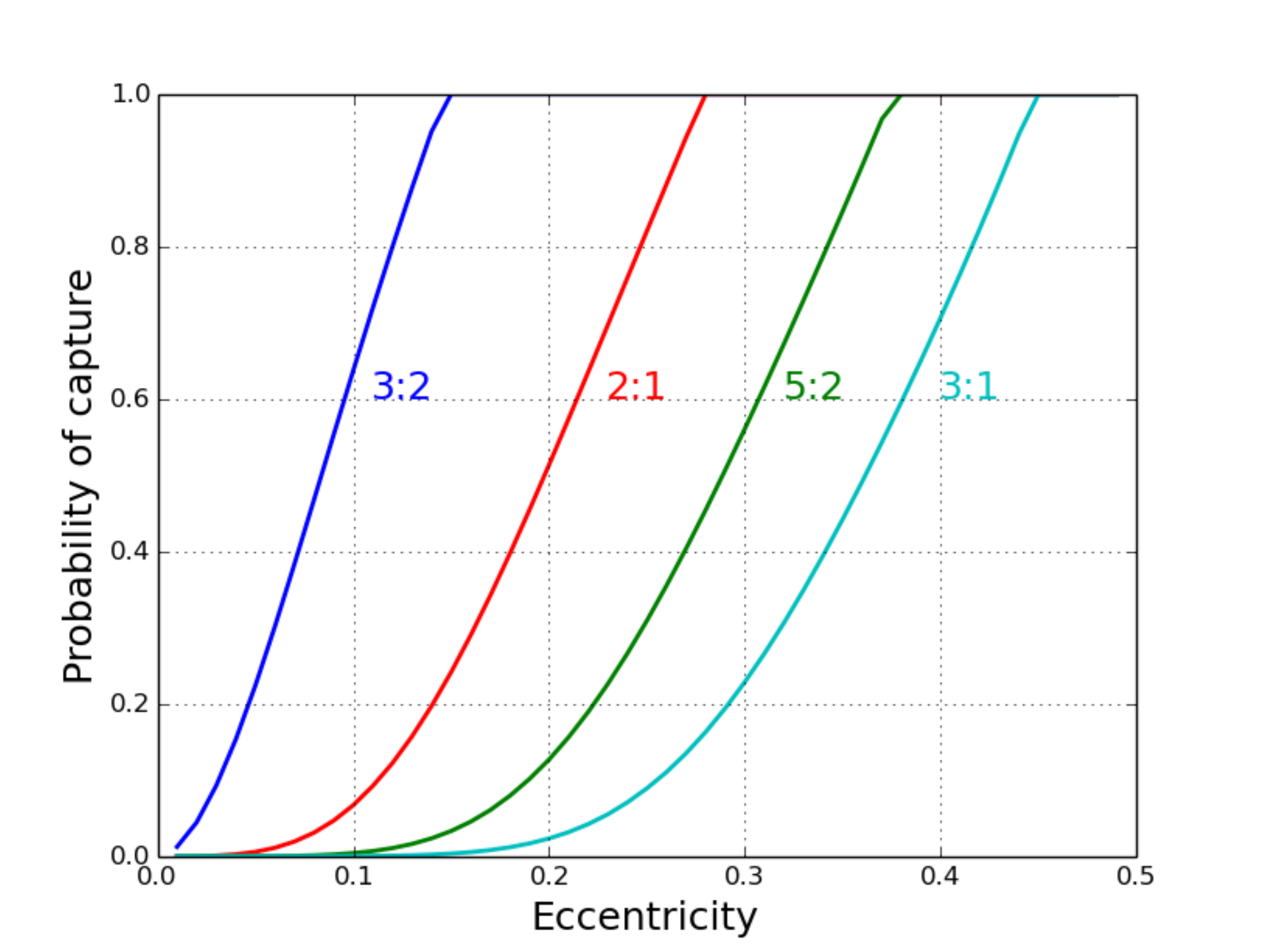}
  \caption{Probability of an isolated resonance capture depending on
    the eccentricity for an El planet $b$. We use the expression of probability by
    \citet{Goldreich66}.}
  \label{fig:eccvsres}
\end{figure}

\section{Discussion and Conclusions}
\label{sec:Conc}

We have produced an updated dynamical study of the GJ667C planetary
system.  We have verified that the spin-orbit evolution of the planets
is exclusively secular.  This result agrees with that of
\citet{Anglada13}. In our tidal model, we also computed the energy
dissipated by tidal friction and the possible associated variation of
the orbital elements. However, we found that the semi-major axis and
eccentricities are not significantly modified by tidal interactions in
this particular system.
 
We have calculated the spin-down times until it reaches a low resonance for
a 6-planet configuration. For this specific configuration and for the 
reported age of the syetem, we found that the equilibrium spin-rotation 
has probably been achieved for planets $b$, $c$, $e$ and $f$ in a 
low resonance. Planet $g$ is probably rotating close to its original 
rate, in the case of planet $d$ it may not have achieved any final 
spin-orbit resonance, this depending on the actual age of the system.

We calculated the capture probabilities of isolated resonances using three
different configurations of gravitational perturbations in the
planetary system: star-planet, star-three planets and star-six
planets.  In agreement to the work by \citet{Anglada13}, we used a 6
compact planetary system. On the other hand, and taking into account
the work by \citet{Feroz14}, we used two other configurations including
2 and 3 planets.

\subsection{Perturbations of Others Planets}

We computed the secular evolution of the system with 2, 3 and 6 planet 
configurations. If we consider the probable age of the
system $\sim$8 Gyr, the absence of a substantial
variation on the semi-major axis is a clear evidence of the system
stability.

The eccentricities of the planets suffer mainly
secular variations. Planets $b$ and $d$ experience the
largest eccentricity changes (see Section \ref{sec:GJ 667C}). 
In the case of a 6-planets configuration, some of them show strong
correlations by pairs, $d$ and $f$ , and $c$ and $e$. The presence of
other planets produces an eccentricity variation around an average
value (see Table \ref{tab:SystemProperties}) that results in a main
frecuency for all the planets around 17000 yr.

We calculated the probability of resonance for El planets
$b$ and $c$ including the perturbations of the other planets. 
We included planet $d$ in a 3-planets configuration, and $e$, $f$
and $g$ in a 6-planets configuration.  The results, shown in Table
\ref{tab:probabilidad}, let us conclude that the secular perturbations
of other planets did not change the final resonance state.

The results of the numerical experiments, including the tidal
perturbation caused by the other planets, showed us that even being a
close-packed system, the gravitational effect of other members did not
cause any change in the final resonance and the probability
of capture is practically the same. Only in the first case, when we go
from a 1 to a 3-planets configuration, the probability changes for the 
3:2 and 2:1 resonance. In the other case, when we went to the 6-planets
configuration the probability did not change at all.

The presence of more planets only shows a little effect
on the oscillation around the resonance at the end of the
spin-orbit evolution, just before reaching the final resonance,
especially for El planet $c$ (see Figure \ref{fig:perturbed}).

After analysing the dynamical evolution of 2, 3 and 6 planets (see
Section \ref{sec:GJ 667C}), we conclude that this particular
system have a secular evolution. There are no variations on the
semi-major axis caused by the other planets. Variations on
eccentricity are also mainly secular.

We conclude that the gravitational influence of other planets 
is negligible for the spin-down times and for the probabilities of
capture in low resonances.

In respect to the stellar system, it is a triple-star system, where the main 
pair of the system GJ 667A and GJ 667B are 0.73 and 0.69 $M_{\odot}$ 
respectively. The pair is separated by $12.6 \ au$ between them and $230 \ au$ 
from GJ 667C.

With the purpose of figuring out the actual influence of the binary pair 
over the dynamical evolution of the system around star C, we
developed numerical experiments including AB stars and the C star
with planet $c$. The star-planet system orbits the center of mass of
the binary. We let the system evolve during a period higher than the
orbital period of the C star.

Although pair AB, the orbit of 
star C and the orbit of planet $c$ around this one, did not suffer any 
changes at all. Even when the system C-c is close to the periapsis of the orbit,
$\sim 180 AU$, the orbit of the planet did not suffer any changes in
its evolution.

It should be noted that the orbital periods of the planets around star
C are a lot less than the period of the star around the
binary. The period of the C-star is in the order of centuries. On the
other hand, the period of the binary is 42.15 years, far from the main
frequency of the secular variations of the eccentricity, that is close to 17
years.

\subsection{Capture Probabilities in low Resonances}

The most stable resonance is 3:2 for an El composition. This is an 
interesting result if we think about the
distribution of heat around the surface of the planet. There are
recent studies dedicated to the investigation of the photosynthetic potential of planets in 3:2
spin-orbit resonances \citep{Brown14}. The probability of reaching a 3:2 resonance, in 
most cases, appears to be significant.

Composition has an effect on the capture probabilities, decreasing the
probability for resonances 2:1 and 3:2 to zero for a Ww composition. 
Although, this looks contradicts the classical idea, 
due to the dissipation of energy, our results show that Ww planets
dissipated tidal heat faster than the El planets. The result of the
spin-down time can be explained by the larger size of the water-rich 
planet.

\subsection{Habitability of GJ 667C}

Previous works propose the possible location of three 
planets ($c$, $f$ and $e$) or maybe four ($d$) inside the HZ 
\citep{Anglada13}. They suggested that planet
$d$ could be a water-rich world, making it more habitable. Taking into 
account the average mass of the planets and the packed configuration
of the system, they proposed the formation place beyond the snow-line
and a consequent migration to their current positions close to the
star.  This makes the planets in GJ 667C volatile-rich planets
containing substantial amounts of water.

On the other hand, \citet{MB14} dismissed the possible habitability of
planet $c$ based on their results of the spin-orbit evolution and the
tidal heating of the interior, which makes this planet more similar 
to Mercury than Earth. Based on our own results, we can state the following
about the habitability in GJ 667C.

\subsubsection{Planet b} 

This planet is obviously beyond the inner limit of the HZ, this makes
it not habitable at all. The final rotation period reached, 4.8
days. This period places this super-Earth in a \textit{slow rotators}
category according to the classification of \citet{Zuluaga12}. 
Its mass of 5.94 $M_{\oplus}$ is unfauvorable when thinking about 
the possible generation of a protective magnetic field \citep{Zuluaga13}.

\subsubsection{Planet c} 

Its distance to the star places it inside the HZ.  With a final rotation
period of 18.75 days and a mass of 3.86 $M_{\oplus}$, this super-Earth is
classified as a \textit{very slow rotator} \citep{Zuluaga12}. This
means that if planet $c$ had an active magnetic field, the dynamo 
would be shut down by now. The absence of a protective magnetic field 
directly leads to an irreversible volatile loss. If at any point in the 
past, planet $c$ had water on its surface, it no longer exists. This implies
that planet $c$ is no longer a habitable planet, even if it was in
the distant past.\\

If we think about the 6-planets configuration, where most of the
planets in GJ 667C finally reach a low resonance, even those inside
the HZ, they will bear the aggressive activity of their host star
during extended periods of time on their diurnal hemispheres. Their
atmospheres (if they have any) would be eroded, causing the loss of
volatiles, like water, in a time shorter than necessary for life's
chemistry to take place.

Finally, as a sideline conclusion, the difference between the
capture probabilities, giveb by composition, could be useful in a
future as a tool to recognize the actual composition of low mass
exoplanets orbiting M-dwarfs, provided that its rotation is feasible
to be measured. Earth-like planets seem more probable to be in a 3:2
resonance than Water-rich planets.

\section{Acknowledgements}

We thank the referee for the valuable insights and comments, all of 
them have been included in the final version of the text.
We want to thank Sebastian Bustamante for helping us with the
calculation of the interior structure of water-rich planets. We
appreciate the revision of the first version of the text made by
Esteban Silva-Villa. We are very grateful with Michael Efroimsky and
Wade Garret Henning for their useful discussion about some of the
structural issues of this paper. We appreciate the
careful revision of the final version of the manuscript made by
Michael Efroimsky, Julien Frouard and Beno\^it Noyelles, their
suggestions were very useful to us. FACom group is supported by
\textit{Estrategia de Sostenibilidad 2015-2016}, Vicerectoría de
Investigación - UdeA and by CODI-UdeA project: IN634CE. Mario Melita
is supported by CONICET-IAFE, Argentina.

\bibliographystyle{mn2e}
\bibliography{bibliography}

\end{document}